\documentclass[aps,pra,letterpaper,onecolumn,accepted=2022-12-01]{quantumarticle}
\pdfoutput=1
\usepackage[utf8]{inputenc}
\usepackage{amsmath}
\usepackage{amsfonts}
\usepackage{braket}
\usepackage{booktabs}
\usepackage[citecolor=magenta, linkcolor=magenta]{hyperref}
\usepackage{graphicx}
\usepackage[pdftex,dvipsnames]{xcolor}
\newcommand{\acr}[1]{\textsc{\lowercase{#1}}}
\newcommand{\Op}[1]{\ensuremath{\mathsf{\hat{#1}}}}
\newcommand{\PE}{\text{PE}}
\newcommand{\tgt}{\text{tgt}}
\newcommand{\Abs}[1]{\left|#1\right|}
\newcommand{\Norm}[1]{\left\lVert#1\right\rVert}
\newcommand{\norm}{\text{norm}}
\newcommand{\tr}{\mathrm{tr}}

\newcommand{\ii}{\mathrm{i}}
\newcommand{\pLoss}{p_{\text{loss}}}
\renewcommand{\Re}{\mathrm{Re}}
\renewcommand{\Im}{\mathrm{Im}}
\newcommand{\re}{\mathrm{re}}
\newcommand{\im}{\mathrm{im}}

\newcommand{\dd}[0]{\,\text{d}}

\newcommand{\iSWAP}{i\textsc{swap}}
\newcommand{\Real}{\mathbb{R}}

\newcommand{\Liouvillian}{\mathcal{L}}
\newcommand{\logical}{\text{L}}
\newcommand{\Bell}{\text{B}}
\DeclareMathSymbol{\shortminus}{\mathbin}{AMSa}{"39}
\newcommand\textmicrosec{{\textmu}s}
\newcommand\microsec{%
\ifmmode\textnormal{\textmicrosec}%
\else\textmicrosec%
\fi}
\newcommand{\ARL}{DEVCOM Army Research Laboratory, 2800 Powder Mill Road, Adelphi, MD 20783, USA}

\begin{document}

\title[Quantum Optimal Control via Semi-Automatic Differentiation]{Quantum Optimal Control \par via Semi-Automatic Differentiation}

\author{Michael H. Goerz}
\affiliation{\ARL}

\author{Sebastián C. Carrasco}
\affiliation{\ARL}

\author{Vladimir S. Malinovsky}
\affiliation{\ARL}

\begin{abstract}
  We develop a framework of ``semi-automatic differentiation'' that combines existing gradient-based methods of quantum optimal control with automatic differentiation. The approach allows to optimize practically any computable functional and is implemented in two open source Julia packages, \texttt{GRAPE.jl} and \texttt{Krotov.jl}, part of the \texttt{QuantumControl.jl} framework. Our method is based on formally rewriting the optimization functional in terms of propagated states, overlaps with target states, or quantum gates. An analytical application of the chain rule then allows to separate the time propagation and the evaluation of the functional when calculating the gradient. The former can be evaluated with great efficiency via a modified \acr{GRAPE} scheme. The latter is evaluated with automatic differentiation, but with a profoundly reduced complexity compared to the time propagation. Thus, our approach eliminates the prohibitive memory and runtime overhead normally associated with automatic differentiation and facilitates further advancement in quantum control by enabling the direct optimization of non-analytic functionals for quantum information and quantum metrology, especially in open quantum systems. We illustrate and benchmark the use of semi-automatic differentiation for the optimization of perfectly entangling quantum gates on superconducting qubits coupled via a shared transmission line. This includes the first direct optimization of the non-analytic gate concurrence.
\end{abstract}
\

%\date{\today}

\maketitle

%\tableofcontents

\section{Introduction}%

Optimal control is a cornerstone in the development of quantum technologies~\cite{BrifNJP2010,Shapiro2012,KochJPCM2016,SolaAAMOP2018,MorzhinRMS2019,Wilhelm2003.10132,KochEPJQT2022}. Quantum information processing~\cite{NielsenChuang2000}, quantum simulation~\cite{GeorgescuRMP2014}, and quantum sensing~\cite{DegenRMP2017} all rely on the ability to manipulate matter at the fundamental quantum level. In simple cases, analytical solutions to the control problem may be possible~\cite{BoscainPRXQ2021}. In more realistic settings, in particular, when taking into account classical or quantum noise, numerical optimization methods must be used. The most general methods for open-loop pulse-level control, gradient-ascent-pulse-engineering (\acr{GRAPE})~\cite{KhanejaJMR2005,FouquieresJMR2011} and Krotov's method~\cite{KrotovEC1983,KrotovCC1988,KrotovBook,SomloiCP1993,BartanaJCP1997,PalaoPRA2003,ReichJCP2012} can both harness the full range of control possible via arbitrary waveform generators~\cite{RashidinejadIJQE2016} or optical pulse shapers~\cite{WeinerRSI2000}. These methods iteratively update the controls via gradient information, that is, the derivative of the optimization functional with respect to the control parameters. Most generally, the control parameters are the amplitudes of time-dependent control fields, e.g., laser pulses in the control of trapped atoms or ions, and microwave pulses for the control of superconducting circuits.

Traditionally, the required gradients have to be derived analytically. In particular for \acr{GRAPE}, the numerical scheme~\cite{KhanejaJMR2005} is formulated specifically for an overlap with a target state. This limits existing \acr{GRAPE} implementations~\cite{JohanssonCPC2013,MachnesPRA2011,HogbenJMR2011,TosnerJMR2009} to a small class of optimization functionals that include state-to-state transfers and quantum gates. The implementation of Krotov's method~\cite{GoerzSPP2019} provides slightly more flexibility, but still requires that a function to evaluate the gradient is passed along with the optimization functional.

The limitation to standard functionals~\cite{PalaoPRA2003} has restrained the full potential of quantum control. In many applications, both in quantum information and in quantum sensing, there are figures of merit that capture the true objective of the optimization but do not fit into the simple mold of reaching a specific target state. For example, the true intent of a control scheme is often to generate entanglement. Similarly, in metrology, the quantum Fisher information~\cite{BraunsteinPRL1994} directly measures the metrological gain~\cite{PezzePRL2009,MaPR2011}.

In the context of universal quantum computing, in combination with single-qubit gates, \emph{any} perfectly entangling two-qubit gate is sufficient to implement a quantum circuit~\cite{NielsenChuang2000}. For a complex system, which specific gate can be implemented with minimal resources or with maximum noise robustness is not predictable, but can be identified by directly maximizing the entanglement power, as defined by the gate concurrence~\cite{KrausPRA2001}. This is complicated by the fact that the gate concurrence is not an analytic quantity and thus there is no closed-form expression for the gradient. To address this, based on the geometric theory of two-qubit gates in the Weyl chamber~\cite{ZhangPRA2003,WattsE2013}, an alternative functional that has an analytic, albeit complicated, gradient was formulated and demonstrated in Refs.~\cite{WattsPRA2015, GoerzPRA2015}.

A breakthrough in the flexibility of quantum optimal control was made by adopting automatic differentiation (AD) in Refs.~\cite{JirariEL2009,LeungPRA2017, AbdelhafezPRA2019, AbdelhafezPRA2020,SchaeferMLST2020}, with proof-of-concept implementations in Refs.~\cite{quantum-optimal-control,qoc}. Automatic differentiation~\cite{Griewank2008, MargossianWIDM2019} considers the evaluation of the optimization functional as a computational graph of elementary operations, and applies the chain rule to evaluate its derivative. This relies on the realization that at a sufficiently low level, \emph{any} function numerically evaluated by a computer is analytic.

In addition to providing the flexibility to include arbitrary final-time functional and running costs, AD has enabled the optimization of open quantum systems through quantum trajectories~\cite{AbdelhafezPRA2019,SchaeferMLST2021,PropsonPRA2022}, which are otherwise difficult to incorporate analytically~\cite{GoerzQST2018}. The approach of gradient-based optimization through AD has also been taken up in some more comprehensive quantum control software packages~\cite{WittlerPRA2021,Ball2001.04060,qgrad}.

Although first developed in the 1960s~\cite{GriewankISMP2012}, the widespread use of AD is associated with the rise of machine learning. It is at the core of the backpropagation method~\cite{RumelhartN1986} for training neural networks. Hence, it is most commonly embedded in machine learning frameworks such as Tensorflow~\cite{Tensorflow}, PyTorch~\cite{PaszkeNIPS2019,pytorch}, \acr{JAX}~\cite{FrostigSYSML2018, jax}, or Flux/Zygote~\cite{Innes1811.01457, InnesJOSS2018,Innes1810.07951, Zygote}. These frameworks have traditionally focused only on the numerical operations required for neural networks, specifically dense real-valued matrix-vector multiplications. The in-place sparse complex-valued linear algebra operations required for the efficient numerical methods of quantum dynamics~\cite{Tal-EzerJCP1984,KosloffJCP1988,BermanJPA1992,KosloffARPC94,AshkenaziJCP1995} have only recently started to be addressed and thus have required workarounds that have hindered performance.

More fundamentally, AD requires the storage of gradient-information for every node in the computational graph. In a ``full-AD'' mode as in Refs.~\cite{JirariEL2009,LeungPRA2017, AbdelhafezPRA2019, AbdelhafezPRA2020,SchaeferMLST2020}, where the full time propagation as well as the evaluation of the optimization functional are performed within the AD framework, this generally implies the storage of a gradient matrix or vector for every linear algebra operation. For large Hilbert space dimensions, open quantum systems, or a large number of control parameters (time steps), this numerical overhead quickly becomes prohibitive.

In this paper, we address both of these issues by introducing the approach of semi-automatic differentiation. The approach exploits the fact that all pulse-level quantum control problems are based on the time evolution of quantum states. We show how formally rewriting the optimization functional as a function of the propagated states, of the overlaps with target states, or of a quantum gate allows to split the evaluation of the gradient into two parts. The main part can be evaluated analytically and leads to an efficient modified \acr{GRAPE} scheme. The remaining part has a profoundly reduced computational complexity and can be evaluated using automatic differentiation with negligible numerical overhead.

The modified \acr{GRAPE} scheme is described in detail and compared with Krotov's method. Thus, the paper gives a blueprint for the efficient implementation of gradient-based optimal control for arbitrary functionals. In addition, we have also implemented the method in the Julia programming language~\cite{BezansonSIREV2017, Julia} in two packages \texttt{GRAPE.jl}~\cite{GRAPE-jl} and \texttt{Krotov.jl}~\cite{Krotov-jl}, both of which are part of a more comprehensive \texttt{QuantumControl.jl} framework~\cite{QuantumControl-jl}.

To demonstrate the numerical efficacy of the semi-automatic differentiation approach, we benchmark the optimization of entangling quantum gates for two superconducting transmon qubits~\cite{JKochPRA2007} with a shared transmission line~\cite{BlaisPRA2007} for a varying Hilbert space dimension and a varying number of time steps. In addition to replicating the results of Refs.~\cite{WattsPRA2015, GoerzPRA2015} with an automatic gradient, we also demonstrate the first \emph{direct} optimization of a non-analytic entanglement measure~\cite{KrausPRA2001}. We show that the numerical cost of the optimization in terms of both memory and runtime scales identically to the direct optimization of a specific quantum gate with an analytic gradient. This is in stark contrast to the use of full automatic differentiation along the lines of Refs.~\cite{JirariEL2009,LeungPRA2017, AbdelhafezPRA2019, AbdelhafezPRA2020,SchaeferMLST2020}, which we show to quickly become infeasible in terms of memory usage and/or runtime.

The paper is structured as follows. In Section~\ref{sec:oct}, we review the efficient implementation of \acr{GRAPE} and Krotov's method. This includes the calculation of gradients to machine precision. The numerical scheme described here also applies to the analytic component of the semi-AD approach, with only minor changes. In Section~\ref{sec:fullad}, we briefly review the concepts of automatic differentiation to develop an understanding of the potential implications of a full-AD approach on numerical costs. Section~\ref{sec:semiad} then develops the theory of semi-automatic differentiation and contains the main new results of this paper. Section~\ref{sec:transmon} defines the optimization problem for perfectly entangling quantum gates on superconducting transmon qubits and shows the benchmarks of the semi-AD and full-AD approaches, as well as the direct optimization of a quantum gate. Section~\ref{sec:conclusion} concludes.

\section{Gradient-based Optimal Control}%
\label{sec:oct}

In this section, we review standard methods of gradient-based numerical control theory. As we will show in section~\ref{sec:semiad}, the method of semi-automatic differentiation that we introduce in this paper builds on these existing control methods, with minimal additional numerical overhead. That is, the numerical cost of optimizing \emph{arbitrary} functionals with semi-automatic differentiation is virtually the same as the cost of optimizing \emph{standard} functionals with traditional gradient-based control methods. Thus, we review the state-of-the-art for implementing these methods as efficiently as possible.

\subsection{Optimization Functionals}

Mathematically, quantum control problems are solved by iteratively minimizing an optimization functional of the general form
\begin{equation}%
  \label{eq:functional}
 J(\{\epsilon_{nl}\})
  = J_T(\{\ket{\Psi_k(T)}\})
  + \int_0^T \!\! g_a\big(\{\epsilon_l(t)\}, t\big) \dd t
  + \int_0^T \!\! g_b\big(\{\ket{\Psi_k(t)}\}, t\big) \dd t\,.
\end{equation}

The terms that constitute the total functional are the final time functional $J_T$ and the running costs $g_{a,b}(t)$ that may encode penalties on the pulse amplitudes or on the propagated states. $J_T$ depends explicitly on the states $\{\ket{\Psi_k(T)}\}$.  These are the result of a forward propagation of some set of states $\{\ket{\phi_k}\}$ at $t=0$ under the control fields $\epsilon_l(t)$. The index $l$ numbers independent control fields. In the example in Section~\ref{sec:transmon}, this would be the real and imaginary part of a complex-valued microwave field in a rotating frame, or equivalently the amplitude and phase of the microwave field in the non-rotating frame. The index $k$ numbers different ``objectives'' that must be achieved simultaneously. For the example of a two-qubit gate $\Op{O}$, $k$ numbers the four logical basis states $\ket{00}$, $\ket{01}$, $\ket{10}$, and $\ket{11}$. Then, a typical functional is
\begin{equation}%
  \label{eq:J_T_sm}
  J_T
  = J_{T, \text{sm}}
  = 1 - \Big\vert\frac{1}{N}
      \sum_{k=1}^N \braket{\phi_k^{\tgt}| \Op{U} | \phi_k}
    \Big\vert^2\,,
\end{equation}
with $N=4$, where $\ket{\phi_k^{\tgt}} \equiv \Op{O} \ket{\phi_k}$ is a target state at $t=T$ and \Op{U} is the time evolution operator, so that $\ket{\Psi_k(T)} \equiv \Op{U} \ket{\phi_k}$.

On the left-hand side the functional $J$ explicitly depends on a set of control parameters $\{\epsilon_{nl}\}$. We will primarily focus here on piecewise constant control fields. In this case, $\epsilon_{nl}$ is the value of the $l$'th control field on the $n$'th interval of the time grid. With $N_T$ time intervals, the time evolution operator is then $\Op{U} = \prod_{n=N_T}^{n=1} e^{-\ii \Op{H}_n dt_n}$ where $\Op{H}_n$ is the Hamiltonian for the $n$'th interval on the time grid, $dt_n = t_{n} - t_{n-1}$, typically
$\Op{H}_n = \Op{H}^{(0)} + \sum_l \epsilon_{nl} \Op{H}^{(l)}$
with the drift Hamiltonian $\Op{H}^{(0)}$ and the control Hamiltonians $\{\Op{H}^{(l)}\}$; although a nonlinear dependency on the control fields is also possible.

While we have written the functional and the time evolution in terms of Hilbert space states and Hamiltonians, all of the formalism applies equally to open quantum systems. In this case, any state $\ket{\Psi}$ is replaced by a density matrix $\Op{\rho}$ and any Hamiltonian $\Op{H}$ is replaced by a Liouvillian super-operator $\Liouvillian$. Any inner product $\braket{\Psi|\Phi}$ is defined for (density) matrices as $\tr[\Op{A}^\dagger \Op{B}]$. In any case, from a numerical perspective, a state $\ket{\Psi}$ or a (vectorized) density matrix $\Op{\rho}$ is a complex vector, and a Hamiltonian $\Op{H}$ or a Liouvillian $\Liouvillian$ is a (sparse) matrix acting on that vector. The main distinction is that $\Op{H}$ for a normal Schrödinger equation is Hermitian, while $\Liouvillian$ is not. For open quantum systems described by a master equation in Lindblad form, the explicit $\Liouvillian$ can be constructed as a sparse matrix from the Lindblad operators; see, e.g., Appendix B.2 in Ref.~\cite{GoerzNJP2014arxiv2}.

\subsection{GRAPE}

The most direct gradient-based method for minimizing a functional as in Eq.~\eqref{eq:functional} is Gradient Ascent Pulse Engineering (\acr{GRAPE})~\cite{KhanejaJMR2005}. In its original form, the optimization is performed by calculating the gradient $\nabla J$ for the piecewise-constant pulse values as the control parameters, i.e., the vector of values $\partial J/\partial \epsilon^{(i)}_{nl}$ for all values $n$, $l$ of the control field in iteration $(i)$, and then update the control field in the direction of the gradient as $\epsilon^{(i+1)}_{nl} = \epsilon^{(i+1)}_{nl} - \alpha (\nabla J)_{nl}$ with some fixed step width (or ``learning rate'') $\alpha$.

In practice, once the gradient $\nabla J$ has been calculated, it can be fed into a black-box gradient-based optimization package,
e.g. the Matlab or SciPy optimization toolboxes~\cite{MatlabOTB,VirtanenNM2020,SciPy} or standalone packages such as \texttt{Optim.jl}~\cite{MogensenJOSS2018}. For one, these packages will perform a linesearch to determine a suitable step width $\alpha$ in each iteration. Even more importantly, they can employ quasi-Newton methods for the optimization that use a Hessian (the matrix of second order derivatives) estimated only from the gradient information of previous iterations. Using second-order information in this way dramatically speeds up convergence~\cite{FouquieresJMR2011} and is strongly recommended.

We find that \acr{L-BFGS-B}~\cite{ByrdSJSC1995}, written in Fortran~\cite{ZhuATMS97} with wrappers for Python~\cite{SciPy} and Julia~\cite{LBFGSB-jl} is a particularly robust quasi-Newton optimizer. It also includes the possibility to apply bounds to the control field (hence the suffix -B). Its particular line search method requires recalculating $\nabla J$ for each step, so it is numerically more expensive than some other linesearch methods that use only \emph{evaluations} of $J$ to determine $\alpha$. Other implementations of \acr{LBFGS}~\cite{MogensenJOSS2018} allow for a variety of linesearch algorithms~\cite{Linesearches-jl} with more customization. However, we have not found any of these to reliably yield better convergence than the method built in to \acr{L-BFGS-B}, and indeed have found the lack of hyperparameters in \acr{L-BFGS-B} to be a virtue.

\subsection{Efficient Evaluation of Final-Time Gradients}%
\label{sec:grad_eval}

Typically, implementations of \acr{GRAPE} only consider specific final-time functionals, $J \equiv J_T$ in Eq.~\eqref{eq:functional}, e.g., $J_T = J_{T,\text{sm}}$ given by Eq.~\eqref{eq:J_T_sm}. We will discuss the efficient evaluation of the gradient for this special case here, and the more general case with non-zero running costs in Section~\ref{sec:running_costs}. To evaluate $\nabla J_T$, we define $\tau_k \equiv \braket{\phi_k^{\tgt} | \Psi_k(T)}$ with $\ket{\Psi_k(T)} = \Op{U} \ket{\phi_k} = \Op{U}_{NT}\dots\Op{U}_1 \ket{\phi_k}$ for each of the $k$ objectives. The time evolution operators $\Op{U}_n$ for the time intervals $n=1\dots N_T$ are piecewise constant. We first consider
\begin{equation}%
  \label{eq:gradtau}
  \nabla\tau^{(k)}_{nl}
  \equiv
  \frac{\partial}{\partial \epsilon_{nl}} \braket{\phi_k^{\tgt} | \Psi_k(T)}
  = \frac{\partial}{\partial \epsilon_{nl}}
    \braket{\phi_k^{\tgt} | \Op{U}_{N_T} \dots \Op{U}_n \dots \Op{U}_1 | \phi_k}
  = \bigg\langle \chi_k(t_{n}) \bigg\vert \frac{\partial \Op{U}_n}{\partial \epsilon_{nl}} \bigg\vert \Psi_k(t_{n-1}) \bigg\rangle
\end{equation}
with $\ket{\chi_k(t_{n})} = U_{n+1}^\dagger \dots U_{N_T}^\dagger \ket{\phi_k^{\tgt}}$, i.e., a backward-propagation of the target state with the adjoint Hamiltonian or Liouvillian and $\ket{\Psi_k(t_{n-1})} = \Op{U}_{n-1}\dots \Op{U}_1 \ket{\phi_k}$, i.e., a forward-propagation of the initial state.

The derivative of the time evolution operator $\Op{U}_n$ of a particular time step acting on an arbitrary state can be evaluated efficiently and to machine precision by defining a ``gradient generator'' for the $n$'th time step as a block matrix
\begin{equation}%
  \label{eq:gradgen}
  G_n = \begin{pmatrix}
    \Op{H}_n & 0 & \dots & 0 &\Op{H}_n^{(1)} \\
    0 & \Op{H}_n & \dots & 0 & \Op{H}_n^{(2)} \\
    \vdots & & \ddots & & \vdots \\
    0 & 0 & \dots & \Op{H}_n & \Op{H}_n^{(L)} \\
    0 & 0 & \dots & 0 & \Op{H}_n
  \end{pmatrix}
\end{equation}
where $\Op{H}_n^{(l)} \equiv \frac{\partial \Op{H}_n}{\partial \epsilon_{nl}}$ for the different controls numbered 1 through $L$, and $\Op{H}_n$ is the full Hamiltonian or Liouvillian for that time step. It can be shown that~\cite{VanLoanITAC1978,GoodwinJCP2015}
\begin{equation}%
  \label{eq:gradprop}
  e^{-\ii G_n dt_n} \begin{pmatrix} 0 \\ \vdots \\ 0 \\ \ket{\Psi} \end{pmatrix}
    =  \begin{pmatrix}
      \frac{\partial}{\partial \epsilon_{n1}} e^{-\ii \Op{H}_n^{(1)} dt_n} \ket{\Psi} \\
      \vdots\\
      \frac{\partial}{\partial \epsilon_{nL}} e^{-\ii \Op{H}_n^{(L)} dt_n} \ket{\Psi} \\
      e^{-\ii \Op{H}_n dt_n} \ket{\Psi}
    \end{pmatrix}
    =  \begin{pmatrix}
      \frac{\partial \Op{U}_n}{\partial \epsilon_{n1}} \ket{\Psi} \\
      \vdots\\
      \frac{\partial \Op{U}_n}{\partial \epsilon_{nL}} \ket{\Psi} \\
      \Op{U}_n \ket{\Psi}
    \end{pmatrix}\,.
\end{equation}
That is, by propagating an extended vector $\ket{\tilde\Psi} = [\ket{\tilde\psi_{1}}, \dots, \ket{\tilde\psi_{L}}, \ket{\Psi}]^T$ under $G_n$, where the components $\ket{\tilde\psi_{l}}$ for the $L$ different controls are initialized to zero, we obtain both the forward propagation of an arbitrary state vector $\ket{\Psi}$ and the gradient of the time evolution operator for that step with respect to every control field. Note that the block matrix defined in Eq.~\eqref{eq:gradgen} does not need to be instantiated; it is sufficient to define it as an abstract operator such that
\begin{equation}%
  \label{eq:apply_gradgen}
  G_n \ket{\tilde\Psi} = \begin{pmatrix}
    \Op{H}_n \ket{\tilde\psi_{1}} + \Op{H}_n^{(1)} \ket{\Psi} \\
    \vdots \\
    \Op{H}_n \ket{\tilde\psi_{L}} + \Op{H}_n^{(L)} \ket{\Psi} \\
    \Op{H}_n \ket{\Psi}
  \end{pmatrix}\,,
\end{equation}
where $\ket{\tilde\psi_{l}}$ is the $l$'th block of the extended $\ket{\tilde \Psi}$ and $\ket{\Psi}$ is the state vector on which $\ket{\tilde \Psi}$ is based.
Thus, the data structure encoding $G_n$ can be essentially the same as the data structure encoding $\Op{H}_n$.

The propagation defined in Eq.~\eqref{eq:gradprop} can be performed using any propagation method used to evaluate the ``normal'' propagation $\Op{U}_n \ket{\Psi}$; that is, a generic \acr{ODE} solver, or preferably a more efficient polynomial expansion of the time evolution operator $\Op{U}_n = e^{-\ii \Op{H} dt}$ for a uniform time step $dt$. For a Hermitian $\Op{H}$, Chebychev polynomials are the fastest converging polynomial expansion~\cite{GilSeguraTemme2007}.
For the standard Schrödinger equation, propagation with Chebychev polynomials $P_m(x)$ is very straightforward and efficient to implement. The time evolution of a state $\ket{\Psi}$ by a single time step $dt$ for the $n$'th interval of the time grid can be written as
\begin{equation}%
  \label{eq:cheby}
  \Op{U}_n \ket{\Psi}
  = \sum_{m} a_m P_m(-\ii \Op{H}_{n,\norm}) \Ket{\Psi}
  = \sum_{m} a_m \ket{\Phi_m}\,,
\end{equation}
where $\Op{H}_{n,\norm}$  is the Hamiltonian $\Op{H}_n$ normalized to a spectral range within $[-1, 1]$, and using the recursive definition of the Chebychev polynomials,
\begin{subequations}
  \label{eq:cheby_recursion}
  \begin{align}
    \ket{\Phi_0} &= \ket{\Psi}\,,\\
    \ket{\Phi_1} &= -\ii \Op{H}_{n,\norm}\ket{\Phi_0}\,,\\
    \ket{\Phi_m} &= -2\ii\Op{H}_{n,\norm}\ket{\Phi_{m-1}} + \ket{\Phi_{m-2}}\,.
  \end{align}
\end{subequations}
The expansion coefficients $a_m$ for a given spectral range and a uniform time step $dt$ can be calculated analytically and are proportional to Bessel functions~\cite{Tal-EzerJCP1984}. For a non-Hermitian $\Op{H}$ (e.g., a Liouvillian), an expansion into Newton polynomials is suitable~\cite{BermanJPA1992,AshkenaziJCP1995,Tal-EzerSJSC2007}. Both Chebychev and Newton propagation have been implemented in Julia~\cite{QuantumPropagators-jl}.

In this context, it is worth noting that the eigenvalues of $G_n$ in Eq.~\eqref{eq:gradgen} are the same as the eigenvalues of the underlying $\Op{H}_n$, with an additional $(L+1)$ degeneracy. In particular, for a Hermitian $\Op{H}_n$ the eigenvalues of $G_n$ are real, despite $G_n$ as a whole not being Hermitian. Thus, if $e^{-\ii \Op{H}_n dt}$ can be evaluated via a Chebychev expansion, so can $e^{-\ii G_n dt}$, with the same expansion coefficients.

\begin{figure}[tb]
  \centering
  \includegraphics{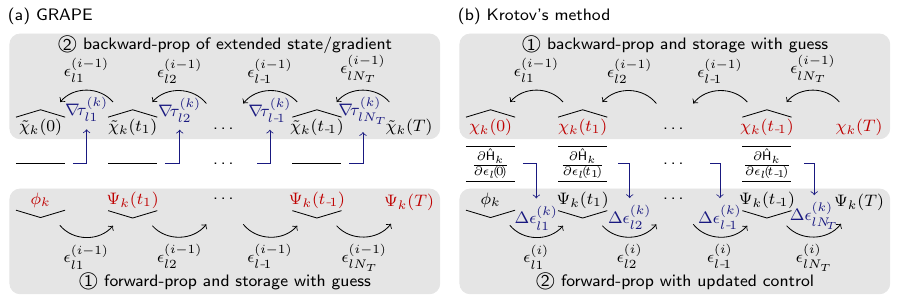}
  \caption{%
    Numerical scheme for the contribution from the $k$'th objective to the gradient/update for the $l$'th control, in iteration $(i)$ of an optimization with GRAPE (a) and Krotov's method (b).
    For GRAPE, the scheme starts in the bottom left with the initial state $\ket{\phi_k}$ at $t=0$. The states marked in red must be stored in memory. The backward propagation of the extended state $\ket{\tilde\chi_k(t)}$ is defined in Eq.~\eqref{eq:gradprop_bw}.
    Both the forward and the backward propagation uses the guess controls, indicated by the superscript $(i-1)$.
    A negative time index, e.g. in $t_{\shortminus\!1}$, is shorthand for $N_T - 1$. The gradient values $\nabla\tau^{(k)}_{nl}$ marked in blue are defined in Eq.~\eqref{eq:gradtau_chi}.
    For Krotov's method, the schemes starts in the top right of panel (b) with $\ket{\chi_k(T)}$ defined in Eq.~\eqref{eq:chi_boundary}.
    The backward propagation uses the guess controls $(i-1)$, while the forward propagation uses the updated controls $(i)$. The updates $\Delta\epsilon_{nl}^{(k)}$ marked in blue correspond to the terms under the sum in Eq.~\eqref{eq:krotov_update}, at the midpoint of the $n$'th time interval. That is, they are summed over $k$ to obtain the total updated $\epsilon_{nl}^{(i)}$.
  }
  \label{fig:schemes}
\end{figure}

The complete numerical scheme for evaluating $\nabla \tau^{(k)}_{nl}$ is shown in Fig.~\ref{fig:schemes}~(a). It starts at the bottom left with the initial state $\ket{\phi_k}$. This state is forward-propagated using the values $\epsilon^{(i-1)}_{nl}$ for the $l$'th control and the $n$'th time step, where the superscript $(i-1)$ indicates the guess for the current iteration $(i)$. The propagation continues to the final state $\ket{\Psi_k(T)}$. All of these propagated states must be stored in memory. After the forward-propagation ends, we initialize an extended state $\ket{\tilde\chi_k(t = t_{N_T} = T)} = [0, \dots 0, \ket{\phi_k^{\tgt}}]^T$,
consisting of a zero block for each of the $L$ controls $\epsilon_l(t)$ and the target state for the objective $(k)$ in the bottom block. This extended state is backward-propagated as $\ket{\tilde\chi_k(t_{n-1})} = e^{-\ii G_{n}^* dt_{n}} \ket{\tilde\chi_k(t_{n})}$ with a negative time step $dt_n$. In the full block-form of the extended state and generator, each propagation step is defined as
\begin{equation}%
  \label{eq:gradprop_bw}
  \begin{pmatrix} \ket{\tilde\chi_{k1}(t_{n-1})} \\ \vdots \\ \ket{\tilde\chi_{kL}(t_{n-1})} \\ \ket{\chi_k(t_{n-1})} \end{pmatrix}
  = \begin{pmatrix}
     \frac{\partial \Op{U}^{\dagger}_n}{\partial \epsilon_{n1}} \ket{\chi_k(t_{n})} \\
      \vdots\\
      \frac{\partial \Op{U}^{\dagger}_n}{\partial \epsilon_{nL}} \ket{\chi_k(t_{n})} \\
      \Op{U}^{\dagger}_n \ket{\chi_k(t_{n})}
    \end{pmatrix}
  = \exp \left[-\ii \begin{pmatrix}
    \Op{H}^\dagger_n & 0 & \dots & 0 &\Op{H}_n^{(1)\dagger} \\
    0 & \Op{H}^\dagger_n & \dots & 0 & \Op{H}_n^{(2)\dagger} \\
    \vdots & & \ddots & & \vdots \\
    0 & 0 & \dots & \Op{H}^\dagger_n & \Op{H}_n^{(L)\dagger} \\
    0 & 0 & \dots & 0 & \Op{H}^\dagger_n\,,
  \end{pmatrix} dt_n\right]
  \begin{pmatrix} 0 \\ \vdots \\ 0 \\ \ket{\chi_k(t_n)} \end{pmatrix}\,,
\end{equation}
which is the backward version of the forward step defined in Eqs.~(\ref{eq:gradgen}--\ref{eq:apply_gradgen}). After each step in the backward propagation, we calculate the $nl$'th component of the gradient of $\tau_k$ with respect to the control values,
\begin{equation}%
  \label{eq:gradtau_chi}
  \nabla\tau_{nl}^{(k)}
  \equiv (\nabla\tau_k)_{nl}
  = \Braket{\tilde\chi_{kl}(t_{n-1}) | \Psi_k(t_{n-1})}\,,
\end{equation}
where $\ket{\tilde\chi_{kl}(t_{n-1})}$ is the state from the $l$'th  block of $\ket{\tilde\chi_k(t_{n-1})}$ and $\ket{\Psi_k(t_{n-1})}$ is read from the stored forward-propagated states. After calculating this overlap, the first $L$ blocks of $\ket{\tilde\chi_k(t_{n-1}}$ must be zeroed out before performing the next step in the backward propagation, so as to have the correct state on the right-hand side of Eq.~\eqref{eq:gradprop_bw}.

The scheme depicted in Fig.~\ref{fig:schemes}~(a) is inherently parallel in the different objectives (index $k$). This is in fact one reason why we have framed the optimization of a quantum gate in terms of multiple objectives, one for each of the logical basis states.
The approach is in contrast to the approach taken by most existing \acr{GRAPE} implementations~\cite{MachnesPRA2011, JohanssonCPC2013, LeungPRA2017} that treat the quantum gate $\Op{U}$ as the dynamic object and may even use explicit matrix-exponentiation to calculate $\Op{U} = e^{-\ii \Op{H} dt}$, limiting the implementation to small Hilbert space dimensions.
In addition to numerical parallelizability and efficiency, formulating the control problem in terms of multiple simultaneous objectives also allows the method to extend to gate optimization in open quantum systems~\cite{GoerzNJP2014} or to ensemble optimization for robust quantum gates~\cite{GoerzPRA2014}.

In principle, we could reverse the order of the backward and forward propagation: Instead of the procedure shown in Fig.~\ref{fig:schemes}~(a), we could first backward-propagate $\ket{\chi_k(T)} = \ket{\phi_k^{\tgt}}$ and store the resulting $\ket{\chi_k(t)}$, and then forward-propagate an extended state $\ket{\tilde\Psi_k(t)}$ to evaluate $\nabla\tau_{nl}^{(k)}$, using Eqs.~(\ref{eq:gradgen}--\ref{eq:apply_gradgen}). This corresponds to $\frac{\partial\Op{U}_n}{\partial \epsilon_{nl}}$ in Eq.~\eqref{eq:gradtau} acting to the right instead of to the left. However, as we show in Section~\ref{sec:semiad_chi}, doing the forward-propagation first is necessary when combining \acr{GRAPE} with automatic differentiation.

Having evaluated $\nabla\tau_{nl}^{(k)}$ via Eq.~\eqref{eq:gradtau}, respectively Eq.~\eqref{eq:gradtau_chi}, we find for the gradient of the square-modulus functional in Eq.~\eqref{eq:J_T_sm}
\begin{equation}%
  \label{eq:grad_J_T_sm}
  (\nabla J_{T,\text{sm}})_{nl}
  \equiv \frac{\partial J_{T,\text{sm}}}{\partial \epsilon_{nl}}
  = -\frac{1}{N^2}
    \sum_{k,k'=1}^N \left[
      \frac{\partial \tau_{k'}^*}{\partial \epsilon_{nl}} \tau_k
      + \tau_{k'}^*\frac{\partial \tau_{k}}{\partial \epsilon_{nl}}
      \right]
  = - \frac{2}{N^2} \Re \sum_{k,k'=1}^N
  \tau_{k'}^* \nabla \tau^{(k)}_{nl}\,.
\end{equation}
For other functionals~\cite{PalaoPRA2003}, we would similarly have to compute the chain rule to complete the \acr{GRAPE} scheme.

\subsection{Krotov's method}

As an alternative to \acr{GRAPE}, Krotov's method~\cite{KrotovEC1983,KrotovCC1988,KrotovBook} takes a constructive approach based on time-continuous control fields, $J(\{\epsilon_l(t)\})$ on the left-hand side of Eq.~\eqref{eq:functional}. For a specifically chosen running cost, $g_a(\epsilon_l(t)) = \frac{\lambda_a}{S(t)} [\epsilon_l(t) - \epsilon_l^{\text{ref}}(t)]^2$, with an arbitrary ``update shape'' $S(t)$ and scaling factor $\lambda_a$, and a reference field $\epsilon_l^{\text{ref}}(t)$ that is typically chosen in each iteration $(i+1)$ as the guess field $\epsilon_l^{(i)}(t)$ for that iteration, the derivation of Krotov's method~\cite{SomloiCP1993,BartanaJCP1997,PalaoPRA2003,ReichJCP2012} considers the necessary and sufficient conditions for the functional derivative $\frac{\partial J}{\partial \epsilon_l(t)}$ to ensure monotonic convergence, $J(\{\epsilon_l^{(i+1)}(t)\}) \le J(\{\epsilon_l^{(i)}(t)\})$, and finds a first-order~\cite{ReichJCP2012} update equation for $\Delta \epsilon^{(i)}_l(t) \equiv \epsilon_l^{(i+1)}(t) - \epsilon_l^{(i)}(t)$ in each iteration,
\begin{equation}%
  \label{eq:krotov_update}
  \Delta\epsilon^{(i)}_l(t) = \frac{S(t)}{\lambda_a} \Im\left[\,
    \sum_{k=1}^N\Braket{\chi_k^{(i-1)}(t)|\frac{\partial \Op{H}}{\partial \epsilon_l^{(i)}(t)}|\Psi^{(i)}_k(t)}
    \right]\,,
\end{equation}
where $\ket{\Psi_k(t)}$ is the initial state $\ket{\phi_k}$ forward-propagated with the updated pulse for the current iteration, and $\ket{\chi^{(i-1)}_k(t)}$ is a state backward-propagated according to
\begin{equation}%
  \label{eq:krotov_bw_eqm}
  \frac{\partial}{\partial t} \ket{\chi_k^{(i-1)}(t)}
  = -\ii \Op{H}^\dagger \ket{\chi_k^{(i-1)}(t)} +
    \frac{\partial g_b}{\partial \bra{\Psi_k^{(i-1)}(t)}}
\end{equation}
with the boundary condition
\begin{equation}%
  \label{eq:chi_boundary}
  \ket{\chi_k^{(i-1)}(T)}
  = - \frac{\partial J_T}{\partial \bra{\Psi_k^{(i-1)}(T)}}\,.
\end{equation}
The discretization to a time grid is done only after formulating the time-continuous update equation and results in the scheme shown in Fig.~\ref{fig:schemes}~(b). For each objective indexed by $k$, we start in the top right with $\ket{\chi_k(T)}$ defined by Eq.~\eqref{eq:chi_boundary}. This state is backward-propagated as
\begin{equation}
  \ket{\chi_k(t_{n-1})}
  = e^{-\ii \Op{H}_{n}^\dagger dt_{n}} \ket{\chi_k(t_{n})}
\end{equation}
($dt_n < 0$) if there are no state-dependent running costs ($g_b \equiv 0$), or using an inhomogeneous propagator otherwise~\cite{NdongJCP2009}. All backward-propagated states must be stored in memory. For each control field indexed by $l$, the control update for the first time step is calculated according to Eq.~\eqref{eq:krotov_update} for $t=0$, i.e., using the result of the backward propagation and the initial state $\ket{\phi_k}$ for each objective. The updated controls for the first time step then allow to obtain $\ket{\Psi_k(t_1)}$, which together with $\ket{\chi_k(t_1)}$ from the stored backward-propagated states allows to calculate the update for the second time step. In this sense, the scheme is \emph{sequential}: the update in every time step depends directly on the state forward-propagated under the updated pulse from the previous time step. In contrast, the update in \acr{GRAPE} is \emph{concurrent}, with independent $\nabla\tau_{nl}^{(k)}$. Like the \acr{GRAPE} scheme in panel~(a), Krotov's method is inherently parallel in $k$, with the caveat that the parallelization must be synchronized after each time step in the forward propagation, so as to evaluate the sum over $k$ in Eq.~\eqref{eq:krotov_update}.

\section{Automatic Differentiation}%
\label{sec:fullad}

Existing implementations of \acr{GRAPE}~\cite{JohanssonCPC2013,MachnesPRA2011,HogbenJMR2011,TosnerJMR2009,WittlerPRA2021} typically hard-code the optimization functional to something equivalent to Eq.~\eqref{eq:J_T_sm}. In order to extend to more functionals, at best a user has to manually supply a routine that calculates the gradient, for example, with a chain rule such as Eq.~\eqref{eq:grad_J_T_sm}. In the worst case, if the gradient is not easily expressed as overlaps $\tau_k$, the numerical scheme in Fig.~\ref{fig:schemes}~(a) would have to be adapted to that particular functional.

Even more problematic are figures of merit that are highly relevant to quantum engineering but do not have an analytic gradient. Typical examples are entanglement measures~\cite{KrausPRA2001} or the quantum Fisher information~\cite{BraunsteinPRL1994} which is directly connected to metrological gain in quantum sensing applications~\cite{PezzePRL2009, MaPR2011}. Evaluating these measures, in general, involves eigenvalue decompositions, which does not lend itself to an analytical calculation of the gradient. Even if equivalent analytic expressions, e.g., for the gate concurrence, can be found~\cite{WattsPRA2015,GoerzPRA2015}, their analytic gradients are extremely tedious to calculate and implement~\cite{GoerzPhd2015}.

In situations where the \emph{analytic} calculation of a gradient is impossible or impractical, the \emph{numerical} evaluation of gradients can be an alternative, in particular through automatic differentiation (AD)~\cite{JirariEL2009,LeungPRA2017, AbdelhafezPRA2019, AbdelhafezPRA2020,SchaeferMLST2020,Griewank2008, MargossianWIDM2019}. The core of AD is the realization that any numerical computation, even one that is seemingly non-analytical like an eigen-decomposition, can ultimately be expressed in elemental numerical steps -- sums and products of floating point numbers, if taken to the extreme of machine instructions. These elemental steps have a known derivative, and thus the gradient of any function can be evaluated by applying the chain rule ad nauseam. Doing this requires that the computer keep track of all intermediate values in the computation. In our case, the function of interest is the optimization functional $J$, with input values $\{\epsilon_{n}\}$ (we temporarily drop the index $l$ numbering different controls here for simplicity).

Consider
\begin{equation}
  J(\epsilon_1, \epsilon_2) = \sin(\epsilon_1) + \epsilon_1 \sqrt{\epsilon_2}
\end{equation}
as a simple example, borrowed from Ref.~\cite{LeungPRA2017}.
We introduce intermediary values, $v_1 = \epsilon_1$, $v_2 = \epsilon_2$, $v_3 = \sin(v_1)$, $v_4 = \sqrt{v_2}$, $v_5 = v_1 v_4$, $v_6 = v_3 + v_5$, and thus finally $J = v_6$. The full chain rule for the gradient of $J$ is
\begin{equation}
  (\nabla J)_n
  \equiv \frac{\partial J}{\partial \epsilon_i}
  = \frac{\partial J}{\partial v_6}
    \frac{\partial v_6}{\partial v_3}
    \frac{\partial v_3}{\partial v_1}
    \frac{\partial v_1}{\partial \epsilon_n}
    +
    \frac{\partial J}{\partial v_6}
    \frac{\partial v_6}{\partial v_5}
    \frac{\partial v_5}{\partial v_1}
    \frac{\partial v_1}{\partial \epsilon_n}
    +
    \frac{\partial J}{\partial v_6}
    \frac{\partial v_6}{\partial v_5}
    \frac{\partial v_5}{\partial v_4}
    \frac{\partial v_4}{\partial v_2}
    \frac{\partial v_2}{\partial \epsilon_n}\,.
\end{equation}
There are two ways to write the chain rule recursively, allowing to consistently evaluate it numerically for arbitrarily complicated functionals. These correspond to the \emph{forward} and \emph{reverse} modes of automatic differentiation.

For the forward mode, we define the ``tangent'' of an intermediary value as $\dot v_j \equiv {\partial v_j}/{\partial \epsilon_n}$ where we have picked a particular $\epsilon_n$ for which we want to evaluate the derivative. We then find
\begin{equation}%
  \label{eq:tangent}
  \dot v_j = \sum_i \frac{\partial v_j}{\partial v_i} \dot v_i\,,
\end{equation}
where the sum is over all $v_i$ on which $v_j$ depends explicitly. We can evaluate these tangents along with the evaluation of the intermediary values themselves. For the derivative with respect to $\epsilon_1$, in our example, we would have $\dot v_1 = 1$, $\dot v_2 = 0$, $\dot v_3 = \cos(\epsilon_1)$, $\dot v_4 = 0$, $\dot v_5 = \sqrt{\epsilon_2}$, $\dot v_6 = \sqrt{\epsilon_2} + \cos(\epsilon_1)$, which is indeed ${\partial J}/{\partial \epsilon_1}$.
To calculate the full gradient, the entire calculation must be evaluated once for each $\epsilon_{n}$. Thus, calculating the gradient in forward-mode is efficient only if the number of dependent variables is small. On the other hand, for a single dependent variable, the runtime and memory requirements of evaluating the derivative are proportional to those of evaluating $J$ itself. In particular, a tangent $\dot v_j$ does not have to be kept in memory longer than the value $v_j$ itself. This makes forward mode automatic differentiation easy to implement, e.g., by using operator overloading and dual numbers~\cite{Griewank2008}.

For the reverse mode, we define the ``adjoint'' of an intermediary value as $\bar v_j \equiv {\partial J}/{\partial v_j}$. The name ``adjoint'' in this context is unrelated to the Hermitian conjugate in quantum mechanics, denoted by a dagger. The definition of the AD adjoint results in a recursive relationship
\begin{equation}
  \bar v_j = \sum_i \bar v_i \frac{\partial v_i}{v_j}\,,
\end{equation}
where the sum is over all $v_i$ which depend on $v_j$. Note that this is the reverse of Eq.~\eqref{eq:tangent}. Consequently, the adjoints are evaluated backward, starting from $\bar v_6 = {\partial J}/{\partial v_6} = 1$. We then further find $\bar v_5 = 1$, $\bar v_4 = \epsilon_1$, $\bar v_3 = 1$, $\bar v_2 = \epsilon_1/({2\sqrt{\epsilon_2}})$ and $\bar v_1 = \cos(\epsilon_1) + \sqrt{\epsilon_2}$. The final adjoints now contain the derivatives for all of the input parameters. This is the primary benefit of reverse-mode AD: the full gradient can be evaluated at once, making it the preferred method for calculating the gradient of an optimization functional $\Real^{N_T L} \rightarrow \Real$. However, we can only start the calculation of the adjoints once $J$ itself has been evaluated. All intermediary values $v_j$ must be stored together with information on which elemental function was used to obtain the value, as well as all adjoints $\bar v_j$.

In practice, this can be implemented in several ways. Early versions of Tensorflow~\cite{Tensorflow} require that the entire calculation is set up as a computational graph, see Fig.~1 in Ref.~\cite{LeungPRA2017}. A forward-pass through the graph calculates the functional and stores intermediary values, while a backward-pass distributes adjoint information to the parents of each node in the graph. Alternatively, a tabular representation of the graph may be constructed during the forward pass, in what is called a Wengert tape. The backward pass then adds adjoint information to the tape. Lastly, it is possible to transform the source code of a function that evaluates the optimization functional into a new function that first performs the original evaluation, and then inverts the computational steps, splicing in code to calculate the adjoints. These implementation details can have a significant impact on the performance and flexibility; see~\cite{RackauckasWeb2021.12.25} for a discussion of the particular tradeoffs. However, they do not change the fundamental memory requirements associated with reverse-mode AD.

There is considerable flexibility in what is considered an ``elemental'' function for which the adjoint can be defined. Most importantly, linear algebra operations do not have to be handled at the level of scalar operations, alleviating to some extent the overhead associated with AD. Using the rules of matrix calculus~\cite{MagnusNeudecker2019, Petersen2008}, AD adjoints for many operations can be defined~\cite{Giles2008}. This even includes eigen- or \acr{SVD} decompositions~\cite{Giles2008b}.

The more high-level the elemental functions are, the less the numerical overhead of AD\@. However, this comes at the cost of having to define more and more analytical adjoints. This is why many AD frameworks have been slow to adopt operations that are outside of the narrow scope of machine learning, which only requires real-valued dense matrix-vector operations. In contrast, quantum dynamics is inherently described with complex-valued state vectors, and operators are usually sparse. Defining AD adjoints for complex linear algebra operations is possible~\cite{Hjorungnes2011}, but has only recently seen adoption. Another practical issue is that naively, the intermediary values (or vectors) $v_i$ are immutable. Thus, most AD frameworks (including Zygote~\cite{Innes1810.07951, Zygote}, which we have used here) do not support in-place linear algebra (\acr{BLAS}~\cite{BlackfordTMS2002}) operations, which can greatly speed up the simulation of quantum dynamics.

\section{Semi-Automatic Differentiation}%
\label{sec:semiad}

We now develop a method to eliminate the two shortcomings of reverse-mode automatic differentiation: the excessive memory overhead associated with having to store the full computational graph, respectively, a Wengert tape, and the limited support in AD frameworks for the linear algebra operations relevant to simulating the dynamics of a quantum system. We do this by applying an analytic chain rule to the calculation of the gradient. To this end, we introduce intermediary variables $z_j$ and rewrite the functional in terms of these intermediaries,  $J(\{\epsilon_{nl}\}) \rightarrow J(\{z_j(\{\epsilon_{nl}\})\})$. The values $z_j$ may be complex, which requires some care when writing out the chain rule.

In principle, one must separate the $z_j$ into real and imaginary part as independent variables, $J = J(\{\Re[z_j]\}, \{\Im[z_j]\})$, resulting in
\begin{equation}
  \label{eq:grad_zj_real_imag}
  (\nabla J)_{nl}
  \equiv \frac{\partial J}{\partial \epsilon_{nl}}
  = \sum_j \left(
    \frac{\partial J}{\partial \Re[z_j]}
    \frac{\partial \Re[z_j]}{\partial \epsilon_{nl}}
    + \frac{\partial J}{\partial \Im[z_j]}
    \frac{\partial \Im[z_j]}{\partial \epsilon_{nl}}
    \right)\,.
\end{equation}
An elegant alternative is to introduce Wirtinger derivatives,
\begin{align}%
  \label{eq:wirtinger1}
  \frac{\partial J}{\partial z_j}
    &\equiv \frac{1}{2} \left(
      \frac{\partial J}{\partial \Re[z_j]}
      -\ii \frac{\partial J}{\partial \Im[z_j]}
      \right)\,, \\
  \label{eq:wirtinger2}
  \frac{\partial J}{\partial z_j^*}
    &\equiv \frac{1}{2} \left(
      \frac{\partial J}{\partial \Re[z_j]}
      +\ii \frac{\partial J}{\partial \Im[z_j]}
      \right)
    = \left(\frac{\partial J}{\partial z_j}\right)^*\,,
\end{align}
which instead treats $z_j$ and the conjugate value $z_j^*$ as independent variables, so that
\begin{equation}%
  \label{eq:wirtinger_chainrules}
  \frac{\partial J}{\partial \epsilon_{nl}}
  = \sum_j \left(
    \frac{\partial J}{\partial z_j}
    \frac{\partial z_j}{\partial \epsilon_{nl}}
    + \frac{\partial J}{\partial z_j^*}
    \frac{\partial z_j^*}{\partial \epsilon_{nl}}
    \right)
  = 2 \Re \sum_j \frac{\partial J}{\partial z_j}
    \frac{\partial z_j}{\partial \epsilon_{nl}}
    \,.
\end{equation}
The derivative of the complex value $z_j$ with respect to the real value $\epsilon_{nl}$ is defined straightforwardly as
\begin{equation}
  \frac{\partial z_j}{\partial \epsilon_{nl}}
  \equiv
  \frac{\partial \Re[z_j]}{\partial \epsilon_{nl}}
  + \ii \frac{\partial \Im[z_j]}{\partial \epsilon_{nl}}\,.
\end{equation}

Our goal is to choose parameters $z_j$ so that $\partial J/\partial z_j$ can be calculated with automatic differentiation with minimal numerical effort.
That is, we would like the computational graph for $J(\{z_j\})$ to be as small as possible. Additionally, if the number of parameters $\{z_j\}$ can be kept small, forward-mode differentiation or even the use of finite difference may become feasible.
For the second part of the chain rule, we require that $\partial z_j/\partial \epsilon_{nl}$ can be calculated analytically (without the use of AD).

Software frameworks for automatic differentiation such as Zygote~\cite{Zygote} and Tensorflow~\cite{Tensorflow} may define a (mathematically questionable~\cite{Petersen2008}) ``gradient'' of a real-valued function $J$ with respect to a complex vector with elements $z_j$ as
\begin{equation}%
  \label{eq:complex_gradient}
  (\nabla_{z} J)_j
  \equiv \frac{\partial J}{\partial \Re[z_j]} + \ii \frac{\partial J}{\partial \Im[z_j]}\,.
\end{equation}
This differs from the Wirtinger derivatives by a factor of two. Thus,
\begin{equation}
  \label{eq:wirtinger_zygote_grad}
  \frac{\partial J}{\partial z_j} = \frac{1}{2} (\nabla_{z} J)^*_j
\end{equation}
in Eq.~\eqref{eq:wirtinger_chainrules} when using, e.g., Zygote's \texttt{gradient} function for $\nabla_z J$.

\subsection{State functionals}%
\label{sec:semiad_chi}

As a starting point, we can take seriously the explicit dependency of $J_T$ on $\{\ket{\Psi_k(T)}\}$ in Eq.~\eqref{eq:functional} and write the gradient of $J_T$ using the chain rule in the states. To do this, we must combine the Wirtinger derivative with the rules of matrix calculus~\cite{MagnusNeudecker2019, Petersen2008}, and write
\begin{equation}%
  \label{eq:grad_via_chi1}
  (\nabla J_T)_{nl}
  \equiv \frac{\partial J_T}{\partial \epsilon_{nl}}
  = 2 \Re \sum_k \left(
    \frac{\partial J_T}{\partial \ket{\Psi_k(T)}}
    \frac{\partial \ket{\Psi_k(T)}}{\partial \epsilon_{nl}}
  \right)\,.
\end{equation}
With the definition in Eq.~\eqref{eq:wirtinger1}, this corresponds directly to the scalar
\begin{equation}
  \frac{\partial J_T}{\partial \epsilon_{nl}}
  = \sum_{km} \left(
    \frac{\partial J_T}{\partial \Re[\Psi_{km}]}
    \frac{\partial \Re[\Psi_{km}]}{\partial \epsilon_{nl}}
    + \frac{\partial J_T}{\partial \Im[\Psi_{km}]}
    \frac{\partial \Im[\Psi_{km}]}{\partial \epsilon_{nl}}
  \right)\,,
\end{equation}
where $\Psi_{km} = \Braket{m|\Psi_k(T)}$ for any orthonormal basis $\{\ket{m}\}$ corresponds to the $z_j$ in Eq.~\eqref{eq:grad_zj_real_imag}.

In Eq.~\eqref{eq:grad_via_chi1}, we may now recognize that the derivative of the scalar $J_T$ with respect to a column vector $\ket{\Psi_k(T)}$ results, according to the rules of matrix calculus, in a row vector that we may associate with a co-state $\bra{\chi_k}$. Specifically, we may define
\begin{equation}%
  \label{eq:chi_grape}
  \ket{\chi_k(T)} \equiv - \frac{\partial J_T}{\partial \bra{\Psi_k(T)}}
  \quad\Leftrightarrow\quad
  \bra{\chi_k(T)} \equiv - \frac{\partial J_T}{\partial \ket{\Psi_k(T)}}\,,
\end{equation}
with a minus sign that will be motivated in Section~\ref{sec:semiad_krotov}.
Since $\ket{\chi_k(T)}$ does not depend on $\epsilon_{nl}$, we may push it into the derivative and obtain
\begin{equation}%
  \label{eq:grad_via_chi}
  \frac{\partial J_T}{\partial \epsilon_{nl}}
  = - 2 \Re \sum_k \frac{\partial}{\partial \epsilon_{nl}}
    \Braket{\chi_k(T) | \Psi_k(T)}\,.
\end{equation}
We can now see that the term under the sum has the exact same form as Eq.~\eqref{eq:gradtau}, that is, the derivative of a complex overlap of two states, $\tau_k \equiv \Braket{\chi_k(T)|\Psi_k(T)}$, which we know how to evaluate numerically via
Eq.~\eqref{eq:gradtau_chi}, respectively the scheme in Fig.~\ref{fig:schemes}~(a). The only difference is that in the original \acr{GRAPE}, we initialize the extended state $\ket{\tilde\chi_k(T)}$ for the backward propagation from the target state, whereas now we initialize it with Eq.~\eqref{eq:chi_grape}. This also explains why we have chosen to perform the forward-propagation first in Fig.~\ref{fig:schemes}~(a): while in the original \acr{GRAPE}, backward and forward propagation are interchangeable, now we need the result $\ket{\Psi_k(T)}$ of the forward propagation in order to initialize the backward propagation.

We can use automatic differentiation to evaluate Eq.~\eqref{eq:chi_grape} for arbitrary functionals. For example, with Zygote's \texttt{gradient} function to evaluate $\nabla_{\!\Psi_k} J_T$ analogously to Eq.~\eqref{eq:complex_gradient}, we have
\begin{equation}%
  \label{eq:chi_zygote}
  \ket{\chi_k(T)} = -\frac{1}{2} \nabla_{\!\Psi_k} J_T\,,
\end{equation}
where the factor $\frac{1}{2}$ accounts for the difference between the complex gradient and the correct Wirtinger derivative, cf.~Eq.~\eqref{eq:wirtinger_zygote_grad}.

\subsection{Overlap functionals}%
\label{sec:semiad_standard}

When formulating the gradient for the square-modulus functional in Eq.~\eqref{eq:grad_J_T_sm}, we already used the overlap of the forward-propagated state for the $k$'th objective with the target state for that objective, $\tau_k \equiv \Braket{\phi_k^{\tgt}|\Psi_k(T)}$. Many of the most common functionals in quantum control are expressed in terms of overlaps of propagated states with target states~\cite{PalaoPRA2003}. We can exploit this to further analytically simplify the calculation of $\ket{\chi_k(T)}$ in Eq.~\eqref{eq:chi_grape} via automatic differentiation. We find
\begin{equation}
  \label{eq:chi_tau}
  \ket{\chi_k(T)}
  = -\frac{\partial J_T}{\partial \bra{\Psi_k(T)}}
  = - \left(
    \frac{\partial J_T}{\partial \tau_k^*} \;
    \frac{\partial \tau_k^*}{\partial \bra{\Psi_k(T)}}
  \right)
  = -\frac{1}{2} \left(\,\nabla_{\tau_k} J_T \right) \ket{\phi_k^{\tgt}}\,,
\end{equation}
where we have used that only the complex conjugate $\tau_k^* = \Braket{\Psi_k(T)|\phi_k^{\tgt}}$ of the overlap depends explicitly on the co-state $\bra{\Psi_k(T)}$. The gradient $\nabla_{\tau_k} J_T$ defined as in Eq.~\eqref{eq:wirtinger_zygote_grad} can be obtained with automatic differentiation.

We note that for functionals that explicitly depend on overlaps~\cite{PalaoPRA2003}, it is generally not difficult to evaluate the chain rule analytically. Thus, the use of automatic differentiation here is less a matter of necessity than of convenience. It allows us to implement a \acr{GRAPE} optimization package where the user can pass an arbitrary functional $J_T(\{\tau_k\})$ without having to explicitly specify a gradient.

\subsection{Gate functionals}%
\label{sec:semiad_gate}

For the optimization of quantum gates, such as the examples we will explore in Section~\ref{sec:transmon}, it is common to have a logical subspace embedded in a larger physical subspace. The functional $J_T$ in this case can often be written as a function of the achieved gate $\Op{U}_{\logical}$ in the logical subspace.

In this context, $\Op{U}_{\logical}$ is the projection of the full time evolution operator $\Op{U}$ to the logical subspace. Specifically, the entries of the matrix $\Op{U}_{\logical}$ are
\begin{equation}%
  \label{eq:gate_definition}
  (\Op{U}_{\logical})_{ij} = \Braket{\phi_i | \Psi_j(T)}
  \quad \Leftrightarrow \quad
  (\Op{U}_{\logical})^{*}_{ij} = \Braket{\Psi_j(T) | \phi_i}\,,
\end{equation}
where $\{\ket{\phi_i}\}$ are the basis states that span the logical subspace (assumed to be the initial states for the optimization objectives), and each $\ket{\Psi_j(T)}$ is the result of forward-propagating $\ket{\phi_j}$.

We may then calculate the gradient of $J_T$ as in Eqs.~(\ref{eq:chi_grape},~\ref{eq:grad_via_chi}), with a further analytic chain rule,  just as in Section~\ref{sec:semiad_standard} but with the elements $(U_{\logical})_{ij}$ instead of the complex overlaps $\tau_k$:
\begin{equation}%
  \label{eq:chi_gate_proto}
  \ket{\chi_k}
  \equiv - \frac{\partial J_T}{\partial \bra{\Psi_k(T)}}
   = - \sum_{ij}
      \frac{\partial J_T}{\partial\,(U_{\logical})^{*}_{ij}}
      \frac{\partial\,(U_{\logical})^{*}_{ij}}{\partial \bra{\Psi_k(T)}}\,,
\end{equation}
again using the notation of the Wirtinger derivative. We have used that only $(U_{\logical})^{*}_{ij}$ depends explicitly on the co-states $\{\bra{\Psi_k(T)}\}$. Furthermore,
\begin{equation}
  \frac{\partial J_T}{\partial\,(U_{\logical})^{*}_{ij}}
  = \frac{1}{2} (\nabla_{U_{\logical}} J_T)_{ij}
\end{equation}
according to the definitions in Eqs.~(\ref{eq:complex_gradient},~\ref{eq:wirtinger2}), and
\begin{equation}
  \frac{\partial\,(U_{\logical})^{*}_{ij}}{\partial \bra{\Psi_k(T)}}
  = \frac{\partial}{\partial \bra{\Psi_k(T)}} \Braket{\Psi_j(T) | \phi_i}
  = \delta_{jk} \ket{\phi_i}
\end{equation}
with the Kronecker delta $\delta_{jk}$. Thus, Eq.~\eqref{eq:chi_gate_proto} simplifies to
\begin{equation}%
  \label{eq:chi_gate}
  \ket{\chi_k}
  = -\frac{1}{2} \sum_i
    (\nabla_{U_{\logical}} J_T)_{ik} \ket{\phi_i}\,,
\end{equation}
where $\nabla_{U_{\logical}} J_T$ is evaluated via automatic differentiation, e.g. with Zygote's \texttt{gradient} function.

While the simplified Eqs.~(\ref{eq:chi_tau},~\ref{eq:chi_gate}) are not \emph{fundamentally} different from constructing the boundary states $\{\ket{\chi_k}\}$ directly with automatic differentiation, they can help to circumventing numerical instabilities that an AD framework may have for complicated functionals. Also, for very large Hilbert space dimensions, it may eliminate some of the numerical overhead associated with automatic differentiation. While the direct construction of $\ket{\chi_k}$ requires differentiation in a number of variables proportional to the full Hilbert space dimension, using Eq.~\eqref{eq:chi_tau} reduces this to the dimension of the logical subspace, 4 complex numbers in the case of a two-qubit gate. Similarly, for Eq.~\eqref{eq:chi_gate} we have a reduction to the square of the dimension of the logical subspace, that is, 16 complex numbers for a two-qubit gate.

\subsection{Running costs}
\label{sec:running_costs}

So far, we have only discussed the evaluation of gradients for final time functionals $J_T$. We now extend the discussion of semi-automatic differentiation to the running costs $g_{a,b}$ in Eq.~\eqref{eq:functional}. Since we are considering piecewise constant pulses, the integral over the running cost turns into a sum over the time steps. That is, we rewrite Eq.~\eqref{eq:functional} as
\begin{equation}%
  \label{eq:functional_discrete}
  J(\{\epsilon_{nl}\})
  =
  J_T(\{\ket{\Psi_k(T)}\}) +
  \sum_{n=1}^{N_T} \sum_l (g_a)_{nl} +
  \sum_{n=0}^{N_T} (g_b)_{n}\,,
\end{equation}
with
\begin{equation}
  (g_a)_{nl} = \frac{1}{dt_n} g_a(\epsilon_{nl}, dt_n)\,,\qquad
  (g_b)_{n} = \frac{1}{\Delta t_n} g_b(\{\ket{\Psi_k(t_n)}\}, t_n)\,.
\end{equation}
As in Fig.~\ref{fig:schemes}, we define $\ket{\Psi_k(t_n)} = \Op{U}_n \dots \Op{U}_1 \ket{\phi_k}$, $t_0 = 0$, $t_{N_T} = T$, and $\Op{U}_n = \exp[-i \Op{H}_n dt_n]$ as the time evolution operator for the $n$'th time interval, $dt_n = t_{n} - t_{n-1}$. Similarly, $\Delta t_n$ is the time step around the time grid point $t_n$, e.g. $\Delta t_0 = dt_1$, $\Delta t_n = \frac{1}{2}(t_{n+1} - t_{n-1})$ for $1\le n < N_T$, and $\Delta t_{N_T} = dt_{N_T}$. For uniform time grids, $dt_n \equiv \Delta t_n \equiv dt$.

Typically, running costs on the control fields are direct analytic expressions, e.g.,
$g_a(\{\epsilon_{nl}\}) = \lambda_a \epsilon_{nl}^2$ to penalize large amplitudes, with a weight $\lambda_a$. Thus, they are easily included in the gradient, e.g., $(\nabla g_a)_{nl} = 2 \lambda_a \epsilon_{nl}$. For convenience, this can also be done with automatic differentiation. This even extends to penalties on the first and second derivatives of the controls~\cite{LeungPRA2017,AbdelhafezPRA2019,AbdelhafezPRA2020}.

More interesting is the case of state-dependent constraints. Typical examples~\cite{SchirmerNJP2011} include trajectory optimizations,
\begin{equation}
  g_{b,\text{trj}}(\{\ket{\Psi_k(t_n)}\})
  = \lambda_b \sum_k \Norm{\ket{\Psi_k(t_n)} - \ket{\Psi^{\tgt}_k(t_n)}}^2\,,
\end{equation}
where the time evolution of each state $\ket{\Psi_k(t_n)}$ should be close to some target evolution $\ket{\Psi^{\tgt}_k(t_n)}$ with a weight $\lambda_b$, or observable optimizations
\begin{equation}
  g_{b,\Op{D}(t)}(\{\ket{\Psi_k(t_n)}\})
  = \lambda_b \sum_k \Braket{\Psi_k(t_n) | \Op{D}(t_n)| \Psi_k(t_n)}\,,
\end{equation}
where the expectation value of some observable $\Op{D}(t)$ is to be minimized.
A special case of this is the minimization of the population in some forbidden subspace~\cite{PalaoPRA2008}, where $\Op{D}(t_n) \equiv \Op{D}$ is a projector into that subspace.

To obtain the full gradient of a functional with a state-dependent running cost, we apply the same procedure as in Section~\ref{sec:semiad_chi} and find
\begin{subequations}
  \begin{align}
    \frac{\partial J}{\partial \epsilon_{nl}}
    &=
    2 \Re \sum_k \left[
      \frac{\partial J_T}{\partial \ket{\Psi_k(T)}}
      \frac{\partial \ket{\Psi_k(T)}}{\partial \epsilon_{nl}}
      + \sum_{n'=0}^{N_T}
      \frac{\partial\,(g_b)_{n'}}{\partial \ket{\Psi_k(t_{n'})}}
      \frac{\partial \ket{\Psi_k(t_{n'})}}{\partial \epsilon_{nl}}
    \right] \\
    \label{eq:gradJ_rc2}
    &=
    -2 \Re \sum_k \frac{\partial}{\partial \epsilon_{nl}} \left[
      \Braket{\chi_{k}^{(T)} | \Op{U}_{N_T} \dots \Op{U}_1 | \phi_k}
     + \sum_{n'=n}^{N_T}
       \Braket{\xi_{k}(t_{n'}) | \Op{U}_{n'} \dots \Op{U}_1 | \phi_k}
    \right]
  \end{align}
\end{subequations}
with
\begin{equation}%
  \label{eq:chi_boundary_gb1}
  \ket{\chi_k^{(T)}} \equiv - \frac{\partial J_T}{\partial \bra{\Psi_k(T)}}\,,
  \qquad
  \ket{\xi_k(t_{n'})} \equiv - \frac{\partial\,(g_b)_{n'}}{\partial \bra{\Psi_k(t_{n'})}}\,,
\end{equation}
cf.~Eqs.~(\ref{eq:chi_boundary},~\ref{eq:chi_grape}). In the sum over $n'$ in Eq.~\eqref{eq:gradJ_rc2}, we have used that $\ket{\Psi_k(t_{n'})}$ depends on $\epsilon_{nl}$ only for $n' \ge n$. This implies that for the final time interval, $n = N_T$, there is only a single term,
\begin{equation}
    \frac{\partial J}{\partial \epsilon_{N_T l}}
    = -2 \Re \sum_k \bigg\langle
      \chi_k(T) \bigg\vert
      \frac{\partial\Op{U}_{N_T}}{\partial \epsilon_{N_T l}}  \bigg\vert
      \Psi_k(t_{N_T-1})
    \bigg\rangle\,,
\end{equation}
with
\begin{equation}%
  \label{eq:chi_boundary_gb}
  \ket{\chi_k(T)}
  \equiv
  \ket{\chi_k^{(T)}} + \ket{\xi_k(T)}
  =
  - \left(
    \frac{\partial J_T}{\partial \bra{\Psi_k(T)}} +
    \frac{\partial\,(g_b)_{N_T}}{\partial \bra{\Psi_k(T)}}
  \right)
  \,.
\end{equation}
Evaluating the gradient progressively backward in time for $n = (N_T-1) \dots 1$, we then find a recursive relationship
\begin{equation}%
  \label{eq:grape_gb_bw_eqm}
  \frac{\partial J}{\partial \epsilon_{nl}}
  = -2 \Re \sum_k \bigg\langle
    \chi_k(t_n) \bigg\vert
    \frac{\partial\Op{U}_{n}}{\partial \epsilon_{nl}}  \bigg\vert
    \Psi_k(t_{n-1})
  \bigg\rangle\,,
\end{equation}
with
\begin{equation}%
  \label{eq:chi_bw_gb}
  \ket{\chi_k(t_n)} =
    \Op{U}_{n+1}^\dagger \ket{\chi_k(t_{n+1})} -
    \frac{\partial\,(g_b)_{n}}{\partial \bra{\Psi_k(t_n)}}\,.
\end{equation}

Thus, there are no fundamental changes to the scheme in Fig.~\ref{fig:schemes}~(a) in the presence of state-dependent running costs. The states $\{\ket{\phi_k}\}$ must be forward-propagated and stored, and then the extended states $\ket{\tilde\chi_k(t_n)}$ are propagated backward to produce the gradient. The only difference is that the boundary state $\ket{\tilde\chi_k(T)}$ is now constructed based on
Eq.~\eqref{eq:chi_boundary_gb} instead of Eq.~\eqref{eq:chi_grape} (or the target states, in the original \acr{GRAPE}). Furthermore, the backward-propagation uses the discrete inhomogeneous Eq.~\eqref{eq:chi_bw_gb}. The inhomogeneity is calculated using the forward-propagated states stored previously, with the derivative of $g_b$ performed analytically or by automatic differentiation.

\subsection{Semi-AD for Krotov's method and time-continuous schemes}%
\label{sec:semiad_krotov}

We have included a minus sign in the definition of $\ket{\chi_k(T)}$, Eq.~\eqref{eq:chi_grape}, respectively $\ket{\chi_k^{(T)}}$ in Eq.~\eqref{eq:chi_boundary_gb1}, since that results in the exact same definition as the $\ket{\chi_k(T)}$ that is the boundary condition for the backward propagation in Krotov's method, Eq.~\eqref{eq:chi_boundary}.
This allows us to make a strong connection between the most general case of semi-automatic differentiation for \acr{GRAPE} and for Krotov's method.

Evaluating Eq.~\eqref{eq:chi_zygote} with a framework for automatic differentiation like Zygote or Tensorflow is all that is required to bring the concept of semi-automatic differentiation to Krotov's method. In this way, it becomes possible to use Krotov's method to optimize towards any computable functional. Conversely, for functionals where the derivative with respect to the states is known analytically, e.g., because they have already been explored using Krotov's method, that existing code can be shared between an implementation of \acr{GRAPE} and Krotov's method.

We may also observe that the generalization of \acr{GRAPE} and Krotov's method are even more closely related in the time-continuous limit. For $dt \rightarrow 0$, we may use the first-order Taylor expansion of $e^{-\ii \Op{H}_n dt}$ to find
\begin{equation}%
  \label{eq:grape_continuous_limit}
 \nabla J_T \approx
 -2\,dt\, \Im\left[\,
    \sum_{k=1}^N\Braket{\chi_k^{(i-1)}(t)|\frac{\partial \Op{H}}{\partial \epsilon_l(t)}|\Psi^{(i-1)}_k(t)}
    \right]\,,
\end{equation}
with the boundary condition of Eq.~\eqref{eq:chi_grape} for $\ket{\chi_k(T)}$.
With $\Delta\epsilon(t) \propto - \nabla J_T$, this matches Krotov's update equation~\eqref{eq:krotov_update}, up to the concurrent $\ket{\Psi_k^{(i-1)}(t)}$ in Eq.~\eqref{eq:grape_continuous_limit} replacing the sequential $\ket{\Psi_k^{(i)}(t)}$ in Eq.~\eqref{eq:krotov_update}.

Similarly, if there are state-dependent constraints $g_b \not\equiv 0$ in Eq.~\eqref{eq:functional}, the time-continuous limit of Eq.~\eqref{eq:grape_gb_bw_eqm} for the backward propagation in the semi-AD \acr{GRAPE} method corresponds directly to the inhomogeneous Eq.~\eqref{eq:krotov_bw_eqm} for the backward-propagation in Krotov's method. However, the former does not require a genuine inhomogeneous propagator~\cite{NdongJCP2009}: the operator $\Op{U}^{\dagger}_{n+1}$ in Eq.~\eqref{eq:grape_gb_bw_eqm} is a normal time evolution operator, with the inhomogeneity added afterward.

Equation~\eqref{eq:grape_continuous_limit} may provide an alternative implementation of a generalized \acr{GRAPE} scheme in the limit of very small $dt$. The scheme would be nearly identical to Fig.~\ref{fig:schemes}~(b); instead of calculating the pulse update directly, it would calculate the gradient $\nabla J_T$ and feed it into the \acr{L-BFGS-B} optimizer. Considering the time evolution operator only to first order in $dt$ avoids having to construct and propagate the extended state discussed in Section~\ref{sec:grad_eval} and may thus be slightly faster. Finally, Eq.~\eqref{eq:grape_continuous_limit} may also provide a motivation for the exploration of optimization schemes that mix and match sequential and concurrent updates~\cite{SchirmerNJP2011, MachnesPRA2011}.

\subsection{Asymptotic memory usage and checkpointing}
\label{sec:memory}

We can now analyze the asymptotic memory usage of the semi-automatic differentiation procedure in relation to the ``full'' use of automatic differentiation as in Refs.~\cite{JirariEL2009,LeungPRA2017, AbdelhafezPRA2019, AbdelhafezPRA2020,SchaeferMLST2020} in general terms. As discussed in Section~\ref{sec:fullad}, automatic differentiation in general requires the storage of any intermediate value in the evaluation of the functional, that is, the time propagation of the quantum states. Thus, the AD memory overhead depends very much on how exactly this propagation is implemented.

In many of the existing uses of AD for quantum control, the time propagation is achieved by exponentiating the Hamiltonian. The required storage if this exponentiation is performed in a single computational step has been analyzed in Ref.~\cite{Narayanan2022}. The intermediary values in this case are the time evolution operators, i.e., the exponentiated matrices, and the states resulting from the application of those time evolution operators, for every time step. For a Hilbert space dimension of $N_H$, the size of the matrices is $N_H^2$. The required storage for these matrices is thus proportional to $N_T N_H^2$ where $N_T$ is the number of time steps. Asymptotically, this dominates over the storage required for the states, which is proportional to $N_T N_H$.

While matrix exponentiation in a single step minimizes the number of intermediary values in the computation graph and thus the AD memory overhead, the computational complexity for algorithms for matrix exponentiation scales polynomially in matrix-matrix multiplications, which themselves scale quite unfavorably as $N_H^3$. In contrast, expanding the time evolution operator in a polynomial series and applying it directly to the states, e.g., using the Chebychev propagation method outlined in Section~\ref{sec:grad_eval} reduces the computational complexity to something polynomial in matrix-vector products, which scale as $N_H^2$. However, in the context of AD, the polynomial expansion also increases the number of intermediary states. Specifically, for a polynomial of order $M$, there are $M$ intermediary matrices and $M$ intermediary states, and thus the total asymptotic memory overhead increases by a factor of $M$. Typically, the $M$ required for convergence of the polynomial is between 10 and 100, depending on the spectral radius of the Hamiltonian and the size of the time step.

A well-established technique to reduce the excessive memory overhead of automatic differentiation by trading it for an increase in runtime is the use of checkpointing~\cite{MargossianWIDM2019}. The central idea of checkpointing is to store only periodic snapshots of the intermediate variables in the computation graph, and then recalculate the forward pass from the last available checkpoint when calculating the gradient via automatic differentiation. This idea has been applied to the use of automatic differentiation in GRAPE~\cite{Narayanan2022}. A snapshot is taken every $C = \sqrt{N_T}$ time steps in the propagation. The asymptotic memory usage in this case reduces from $N_T N_H^2$ to $(N_C + C) N_H^2 = 2\sqrt{N_T} N_H^2$, where $N_C = N_T/C$ is the number of checkpoints. That is, checkpointing achieves a quadratic reduction with respect to the number of time steps. With a polynomial propagator, we again have an increase by a factor of $M$ within each checkpointed segment of size $C$, thus resulting in a memory scaling of $(N_C + M C) N_H^2 \approx M \sqrt{N_T} N_H^2$, again a quadratic reduction.

The fundamental idea of checkpointing can also be applied in the semi-AD scheme in Fig.~\ref{fig:schemes}~(a). By default, we store all the forward propagated states marked in red. Instead, we may store only every $C$ states. During the backward propagation of $\ket{\tilde\chi_k(T)}$, we then repeat the forward propagation from the last available checkpoint to recover that states $\ket{\Psi_k(t_n)}$ required for the overlaps that determine $\nabla \tau^{(k)}_{ln}$. The required memory for storage is thus reduced from $N_T$ to $N_C + C = 2 \sqrt{N_T}$. However, we need an additional $N_T - N_C = N_T - \sqrt{N_T}$ propagation steps.

Lastly, we can consider the special case of unitary dynamics. No storage at all is required in the semi-AD scheme: after the initial forward propagation in Fig.~\ref{fig:schemes}~(a), the resulting $\ket{\Psi_k(T)}$ can simply be backward-propagated in parallel with $\ket{\tilde\chi_k(T)}$. Thus, we trade the need for storage with a full additional time propagation. The method is not applicable in open quantum systems where the dynamics are not unitary, and thus a backward propagation of $\ket{\Psi_k(T)}$ does not recover the states $\ket{\Psi_k(t_n)}$ from the forward propagation. The unitary dynamics can be exploited in the same way to reduce the storage in a full-AD implementation of GRAPE~\cite{Narayanan2022}.

\begin{table}[tb]
  \centering
  \begin{tabular}{l|cccccc}
    \toprule
                         &  Full-AD      &  Full-AD            &  Full-AD        &  Full-AD              &  Semi-AD    &  Semi-AD           \\
    \midrule[0.25pt]
    prop. meth.          &  exp.         &  exp.               &  polyn.         &  polyn.               &  any        &  any               \\
    \midrule[0.25pt]
    checkpoint           &               &  \checkmark         &                 &  \checkmark           &             &  \checkmark        \\
    \midrule
    matrix ($N_H^2$)     &  $N_T$        &  $N_C + C$          &  $M N_T$        &  $N_C + M C$          &  0          &  0                 \\
    vector ($N_H$)       &  $N_T$        &  $N_C + C$          &  $M N_T$        &  $N_C + M C$          &  $N_T$      &  $N_C + C$         \\
    \midrule
    mem ($\mathcal{O}$)  &  $N_T N^2_H$  &  $(N_C + C) N_H^2$  &  $M N_T N_H^2$  &  $(N_C + M C) N_H^2$  &  $N_T N_H$  &  $(N_C + C) N_H$   \\
    \bottomrule
  \end{tabular}
  \caption{
    Asymptotic memory usage for full automatic differentiation (Full-AD) and semi-automatic differentiation (Semi-AD).
    The table shows how the number of matrices and vectors that must be stored in memory scales with the size of the Hilbert space $N_H$ and the number of time steps $N_T$.
    For full-AD, we compare propagation via matrix exponentiation (``exp.'') and propagation via a polynomial method (``polyn.''), such as the Chebychev polynomials detailed in section~\ref{sec:grad_eval}. In this case $M$ is the order of the polynomial required for convergence, equal to the number of matrix-vector products in a propagation step. Also included is the number of matrices/states that need to be stored when checkpointing is used. $C$ denotes the time steps between checkpoints, and $N_C = N_T / C$ is the number of checkpoints.
    The bottom row shows the total asymptotic scaling of the required memory.
  }
  \label{tab:memory_usage}
\end{table}
The asymptotic memory requirement for both full-AD and semi-AD for different propagators and with and without checkpointing is summarized in Table~\ref{tab:memory_usage}. It is important to point out that for semi-AD, the memory requirement is determined only from the storage of states, and is thus linear both in the Hilbert space dimension and in the number of time steps. The remaining use of automatic differentiation to evaluate $\nabla_{\Psi_k} J_T$ in Eq.~\eqref{eq:chi_zygote}, $\nabla_{\tau_k} J_T$ in Eq.~\eqref{eq:chi_tau}, or $\nabla_{U_L} J_T$ in Eq.~\eqref{eq:chi_gate} is negligible in comparison. At worst, for $\nabla_{\Psi_k}$, it is a constant number of states. At best, for $\nabla_{\tau_k}$, it is a small number of scalars.

In practice, the memory and runtime characteristics of full-AD may be less predictable. Many AD frameworks have large constant overhead and large prefactors for the asymptotic scaling in Table~\ref{tab:memory_usage}. On the other hand, the size of the computational graph and thus the amount of memory required for AD depend heavily on the exact implementation details of the time propagation. Therefore, we will explore the scaling of memory and runtime empirically for a specific implementation and a realistic example in Section~\ref{sec:transmon}.

As discussed in Section~\ref{sec:fullad}, there is a certain freedom to define what operations constitute ``elementary operation'', with a known pre-defined adjoint. For example, at least in principle it would be possible to define an entire propagation step via Chebychev expansion as a single node in the computational graph. This would eliminate the scaling with the number of coefficients $M$, but would require to implement a custom adjoint for that propagation step, which is not trivial. Custom adjoints would have to be implemented by hand for every different propagation method.

In general, any use of automatic differentiation involves a trade-off between memory usage, runtime, and code complexity (with custom adjoints).  We believe that semi-automatic differentiation is a particularly attractive balance of these three goals: It is strictly linear in the number of time steps and the dimension of the Hilbert space, its runtime and structure match that of a traditional GRAPE implementation without any AD capabilities, and it requires no implementation of any custom adjoints, which may otherwise achieve similar memory scaling. Specifically, it works with any propagation method without modification; the runtime of the optimization is directly proportional to the runtime of the time propagation.

\section{Optimizing Gate Concurrence for Two Coupled Transmons}%
\label{sec:transmon}

\subsection{Two-Qubit Gates on Transmons with a Shared Transmission Line}

As an example to benchmark the semi-automatic differentiation approach to optimal control we consider two superconducting transmon qubits~\cite{JKochPRA2007} with a shared transmission line~\cite{BlaisPRA2007} (``cavity'') that allows to control the system via microwave pulses. Each transmon qubit is an anharmonic Duffing oscillator that couples to the transmission line. In the dispersive limit, where the qubit-cavity detuning dominates the coupling strength, the cavity can be eliminated. This results in an effective two-transmon Hamiltonian with a static qubit-qubit coupling. Furthermore, a microwave control field with a frequency $\omega_d$ near the qubit frequencies $\omega_1$, $\omega_2$ drives transitions on the transmons. In the rotating-wave approximation, the effective Hamiltonian reads~\cite{PolettoPRL2012}
\begin{subequations}
  \label{eq:transmon_ham}
  \begin{equation}
    \Op{H}
    = \Op{H}_0
      + \Omega_{\re}(t) \Op{H}_{d,\re}
      + \Omega_{\im}(t) \Op{H}_{d,\im}
  \end{equation}
  with ($\hbar = 1$)
  \begin{align}
    \Op{H}_0
    &=
      \sum_{q=1,2} \left[
        \left(\omega_q - \omega_d + \frac{\alpha_q}{2}\right)
          \Op{b}^{\dagger}_q \Op{b}_q
        - \frac{\alpha_q}{2}(\Op{b}^{\dagger}_q \Op{b}_q)^2
      \right]
      + J \left(
        \Op{b}_1^{\dagger} \Op{b}_2 + \Op{b}_1 \Op{b}_2^{\dagger}
      \right)
    \\
    \Op{H}_{d,\re}
    &=
      \frac{1}{2} \left[
        (\Op{b}_1^{\dagger} + \Op{b}_1)
        + \lambda  (\Op{b}_2^{\dagger} + \Op{b}_2)
      \right]
    \\
    \Op{H}_{d,\im}
    &=
      \frac{\ii}{2} \left[
        (\Op{b}_1^{\dagger} - \Op{b}_1)
        + \lambda  (\Op{b}_2^{\dagger} - \Op{b}_2)
      \right]
  \end{align}
\end{subequations}
where $\Op{b}^{\dagger}_q$ and $\Op{b}_q$ are the creation and annihilation operators for the transmon excitations. Transmon qubits can be engineered to a wide range of frequencies, anharmonicities, and coupling strengths~\cite{GoerzNPJQI2017}. Here, we use the parameters listed in Table~{\ref{tab:transmon_parameters}}, cf.~Ref.~\cite{GoerzPRA2015}, as a typical example.

\begin{table}[tb]
  \centering
  \begin{tabular}{llcrl} \toprule
    left qubit frequency              &  $\omega_1$  &  $=$ & 4.380 & $\cdot \quad 2\pi$\,GHz \\
    right qubit frequency             &  $\omega_2$  &  $=$ & 4.614 & $\cdot \quad 2\pi$\,GHz \\
    rotating frame (drive) frequency  &  $\omega_d$  &  $=$ & 4.498 & $\cdot \quad 2\pi$\,GHz \\
    left qubit anharmonicity          &  $\alpha_1$  &  $=$ &  210  & $\cdot \quad 2\pi$\,MHz \\
    right qubit anharmonicity         &  $\alpha_1$  &  $=$ &  215  & $\cdot \quad 2\pi$\,MHz \\
    effective qubit-qubit coupling    &  $J$         &  $=$ &   -3  & $\cdot \quad 2\pi$\,MHz \\
    relative coupling strength        &  $\lambda$   &  $=$ & 1.03  &         ~\\
  \bottomrule
  \end{tabular}
  \caption{%
    \label{tab:transmon_parameters}
    Parameters for the system of two coupled transmon qubits.
  }
\end{table}

The control field in general is complex-valued, corresponding to variations in both the amplitude and phase relative to the rotating frame; it is easiest to split it into real and imaginary parts and treat them as independent controls $\Omega_{\re}(t)$ and $\Omega_{\im}(t)$, as we have done above.

The Hamiltonian in Eq.~\eqref{eq:transmon_ham} is well-suited as a system on which to benchmark optimal control methods for quantum gates. First, the system is fully controllable, allowing to implement any two-qubit gate if no further restrictions are placed on the control field~\cite{WattsPRA2015, MaSB2021}. Second, entangling quantum gates have been demonstrated with gate durations anywhere between 20~ns and 5~\microsec~\cite{GoerzPRA2015, GoerzNPJQI2017,KjaergaardARCMP2020, BlaisRMP2021}. Thus, we can reasonably explore the numerical scaling of the optimization procedure with the number of time steps. Lastly, the number of levels in the anharmonic oscillator that reasonably contribute to the gate dynamics varies significantly depending on the gate mechanism and the amplitudes and frequencies of the control field~\cite{BlaisRMP2021}. This allows us to benchmark the optimization for varying Hilbert space sizes. Especially for non-analytic controls obtained with optimal control, leakage from the logical subspace can be a significant problem, and we include as many as $N_q=15$ levels in the numerical simulation to account for this.

\subsection{Optimization Functionals}%
\label{sec:functionals}

A primary benefit of the semi-automatic differentiation is that it allows optimizing arbitrary figures of merit, including ones for which it is difficult or impossible to derive analytical gradients. This freedom can significantly enhance the effectiveness of optimal control. For example, in the context of entangling quantum gates, it was demonstrated that optimizing for an arbitrary perfectly entangling quantum gate is easier than optimizing for a \emph{specific} entangling gate such as CNOT~\cite{WattsPRA2015,GoerzPRA2015}. This is because for a given Hamiltonian it is not known a priori \emph{which} perfect entangler will be easiest to implement with given resources such as a maximum power of the control pulse. At the same time, for the realization of a universal quantum computer, one perfect entangler together with arbitrary single-qubit operations is sufficient.

The entangling power of a quantum gate is measured by the gate concurrence, defined as the maximum concurrence that can be obtained by applying the quantum gate to a separable input state~\cite{KrausPRA2001}. It can be evaluated by writing a two-qubit gate $\Op{U}_{\logical}$ (a $4\times4$ matrix in the logical subspace) in a Cartan decomposition~\cite{ZhangPRA2003},
\begin{equation}%
  \label{eq:cartan}
  \Op{U}_{\logical}
  = \Op{k}_1 \exp\left[
    \frac{\ii}{2}(
      c_1 \, \Op{\sigma}_x\Op{\sigma}_x +
      c_2 \, \Op{\sigma}_y\Op{\sigma}_y +
      c_3 \, \Op{\sigma}_z\Op{\sigma}_z
    )
    \right] \Op{k}_2\,,
\end{equation}
where $\Op{\sigma}_{x,y,z}$ are the Pauli matrices, $\Op{k}_{1,2}$ are single-qubit operations, and $c_{1,2,3}$ are real-valued coefficients that characterize the two-qubit aspect of the quantum gate. When eliminating symmetries in Eq.~\eqref{eq:cartan}, the coefficients can be understood as coordinates in a geometric representation of two-qubit gates called the Weyl chamber.  A gate is a perfect entangler with a gate concurrence $C = 1$ if $c_1 + c_2 \ge \frac{\pi}{2}$, $c_1 - c_2 \le \frac{\pi}{2}$, and $c_2 + c_3 \le \frac{\pi}{2}$, which is a polyhedron within the Weyl chamber. Otherwise,
\begin{equation}%
  \label{eq:gate_concurrence}
  C(\Op{U}_{\logical})
    = \max \Abs{\sin(c_{1,2,3} \pm c_{3,1,2})}\,,
\end{equation}
cf.~Ref.~\cite{KrausPRA2001}. The calculation of the Weyl chamber coordinates themselves is described in Ref.~\cite{ChildsPRA2003} and implemented in Refs.~\cite{weylchamber,QuantumControlBase-jl} and involves obtaining the eigenvalues of $\Op{U}_{\logical}$, as well as a branch selection of a complex logarithm. Thus, the evaluation of Eq.~\eqref{eq:gate_concurrence} is inherently non-analytical, preventing the analytic construction of a gradient $\nabla C$ with respect to the control values.

For this reason, a less direct measure for the entangling power of the quantum gate was developed in Refs.~\cite{WattsPRA2015,GoerzPRA2015}. Intuitively, it minimizes the geometric distance from the polyhedron of perfect entanglers in the Weyl chamber. Mathematically, that distance is formulated not in terms of the Weyl chamber coordinates, but in terms of the Makhlin local invariants $g_{1,2,3}$~\cite{MakhlinQIP2002}. Similarly to the Weyl chamber coordinates, these characterize two-qubit gates up to single-qubit operations. It can be shown that the geometric distance to the polyhedron of perfect entanglers is
\begin{equation}%
  \label{eq:D_PE}
  D_{\PE}(\Op{U}_{\logical}) = g_3 \sqrt{g_1^2 + g_2^2} - g_1\,.
\end{equation}
Unlike the Weyl chamber coordinates, the local invariants can be calculated \emph{analytically} from the two-qubit gate $\Op{U}_{\logical}$ as
\begin{equation}
  g_1 = \frac{1}{16} \Re\left[\tr^2(\Op{m})\right]\,, \qquad
  g_2 = \frac{1}{16} \Im\left[\tr^2(\Op{m})\right]\,, \qquad
  g_3 = \frac{1}{4} \left[\tr^2(\Op{m}) - \tr(\Op{m}^2) \right]\,,
\end{equation}
with $\Op{m} = \Op{U}_{\Bell}^T \Op{U}_{\Bell}$, where $\Op{U}_{\Bell}$ is the representation of $\Op{U}_{\logical}$ in the Bell basis. Applying matrix calculus, it is possible -- although both tedious and lengthy -- to calculate an analytic gradient of Eq.~\eqref{eq:D_PE}~\cite{GoerzPhd2015}. For the derivative with respect to a state, cf.~Eqs.~(\ref{eq:chi_boundary},~\ref{eq:chi_grape}), a Python implementation is available in Ref.~\cite{weylchamber}.

Both $C(\Op{U}_{\logical})$ and $D_{\PE}(\Op{U}_{\logical})$ are only well-defined if $\Op{U}_{\logical}$ is unitary.
To ensure this in the optimization, we may add a term that calculates the loss of population from the logical subspace,
\begin{equation}%
  \label{eq:pop_loss}
  \pLoss(\Op{U}_{\logical}) = 1 - \frac{1}{4}\tr\left[\Op{U}_{\logical}^{\dagger}\Op{U}_{\logical}\right]\,.
\end{equation}

Altogether, we use the following three optimization functionals:
\begin{enumerate}
  \item Square-modulus gate optimization (SM)
    \begin{equation}%
      \label{eq:benchmark_J_T_sm}
      J_{T, \text{sm}}(\{\tau_k\})
      = 1 - \Big\vert
        \frac{1}{4} \sum_{k=1}^{4}
        \underbrace{
          \Braket{\phi^{\tgt}_k | \Psi(T)}
        }_{\equiv\tau_k}
      \Big\vert^2
      = 1 - \frac{1}{16} \sum_{k=1}^{4} \sum_{k'=1} ^{4} \tau_{k'}^* \tau_k\,,
    \end{equation}
    cf. Eq.~\eqref{eq:J_T_sm}, where $\ket{\phi_k^{\tgt}}$ is the result of applying the target gate
    \begin{equation}
      \Op{O}
      = \sqrt{\text{\iSWAP}}
      \equiv \begin{pmatrix}
        1  &          0           &          0           &  0 \\
        0  &   \frac{1}{\sqrt{2}} & \frac{\ii}{\sqrt{2}} &  0 \\
        0  & \frac{\ii}{\sqrt{2}} &   \frac{1}{\sqrt{2}} &  0 \\
        0  &          0           &          0           &  1 \\
      \end{pmatrix}
    \end{equation}
    to the initial states $\ket{\phi_1} = \ket{00}$, $\ket{\phi_2} = \ket{01}$, $\ket{\phi_3} = \ket{10}$, $\ket{\phi_4} = \ket{11}$ spanning the logical subspace.
    The choice of $\sqrt{\text{\iSWAP}}$ as the target gate is somewhat arbitrary, although it has been demonstrated that this is an easily reachable gate for the Hamiltonian in Eq.~\eqref{eq:transmon_ham}~\cite{GoerzPRA2015}.
  \item Perfect entangler optimization (PE)
    \begin{equation}%
      \label{eq:benchmark_J_T_PE}
      J_{T, \text{PE}}(\Op{U}_{\logical}) = \frac{1}{2}\left(1-D_{\PE}(\Op{U}_{\logical})\right) + \frac{1}{2} \pLoss(\Op{U}_{\logical})\,,
    \end{equation}
    where $D_{\PE}$ and $\pLoss$ are defined in Eq.~\eqref{eq:D_PE} and Eq.~\eqref{eq:pop_loss}, respectively.
  \item Concurrence optimization (C)
    \begin{equation}%
      \label{eq:benchmark_J_T_C}
      J_{T, \text{C}}(\Op{U}_{\logical}) = \frac{1}{2}\left(1-C(\Op{U}_{\logical})\right) + \frac{1}{2} \pLoss(\Op{U}_{\logical})\,,
    \end{equation}
    where $C$ and $\pLoss$ are defined in Eq.~\eqref{eq:gate_concurrence} and Eq.~\eqref{eq:pop_loss}, respectively. This is an example of a non-analytic functional whose gradient can only be evaluated via automatic differentiation.
\end{enumerate}

\subsection{Benchmarks}

\begin{figure}[p]
  \centering
  \includegraphics{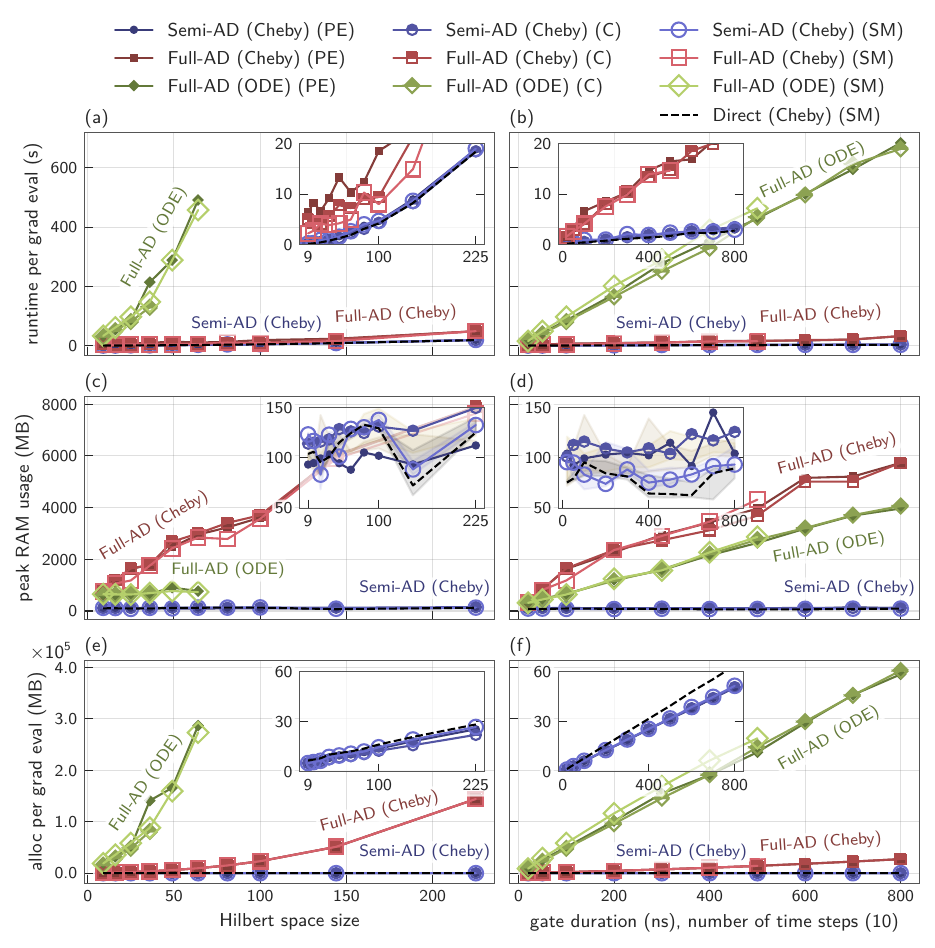}
  \caption{%
    Benchmarks for 10 iterations of a GRAPE optimization of entangling gates on two coupled transmons for different functionals and different usage of automatic differentiation.
    The top row shows the optimization runtime in seconds per evaluation of the gradient (forward propagation of the basis states and backward propagation of the extended gradient-states);
    the center row shows the median peak RAM usage in megabyte from up to 20 optimization runs, where the insets also indicate the range of peak RAM usage as shaded areas; and the bottom row shows the memory allocations per gradient evaluation in megabyte. The left column shows how these quantities vary with the size of the Hilbert space;
    the right column shows the data for different gate durations in nanoseconds. This is directly proportional to the number of time steps ($dt=0.1~\text{ns}$). Each panel shows the performance of an optimization towards a perfect entangler in the Weyl chamber (PE) or by maximizing the gate concurrence (C) directly. Additionally, the data for the optimization of a $\sqrt{\text{iSWAP}}$ gate via a square-modulus (SM) functional is shown. The dynamics are simulated by evaluating the time evolution operator in Chebychev polynomials (``Cheby''), or by using a generic ODE solver.
    The gradient is evaluated either via semi-automatic differentiation (``Semi-AD'') for the PE and C optimizations or via a set over overlaps $\{\tau_k\}$ for the SM optimization, via full automatic differentiation (``Full-AD''), or -- for the ``Direct'' optimization -- using an analytic gradient (no automatic differentiation).
  }
  \label{fig:benchmark}
\end{figure}

The result of benchmarking the functionals in Section~\ref{sec:functionals} with semi-automatic and full automatic differentiation is shown in Fig.~\ref{fig:benchmark}.
The ``Semi-AD'' optimization uses the approach described in Section~\ref{sec:semiad_standard} for the gate optimization with the square-modulus functional (SM), and the approach described in Section~\ref{sec:semiad_gate} for the perfect-entanglers (PE) and concurrence (C) optimizations, as indicated by the argument on the left-hand-sides of Eqs.~(\ref{eq:benchmark_J_T_sm}--\ref{eq:benchmark_J_T_C}). The resulting optimized control fields are similar to those obtained in Ref.~\cite{GoerzPRA2015}, and are shown in the ``PE'' examples in the online documentation of the \texttt{GRAPE.jl} and \texttt{Krotov.jl} packages~\cite{GRAPE-jl, Krotov-jl}, which implement both the direct and semi-AD optimization methods.

All of the optimizations use an expansion in Chebychev polynomials for the time propagation, Eq.~\eqref{eq:cheby}. We compare this with an optimization using full automatic differentiation (``Full-AD''). This means that we simply propagate the set of initial states with an AD-aware \acr{ODE} solver, the Runge-Kutta algorithm \acr{DP5}~\cite{DormandJCAM1980} in \texttt{DifferentialEquations.jl}~\cite{RackauckasJORS2017} and then evaluate the functional within the Zygote AD framework~\cite{Zygote}. As an alternative to the general \acr{ODE} solver, we also benchmark a full-AD variant of the Chebychev propagator. In this implementation, in-place linear algebra operations must be avoided, adding runtime overhead due to the allocation and deallocation of memory, cf. panels~(e, f). It is possible to do this within Zygote without too much numerical overhead due to the simplicity of Eqs.~(\ref{eq:cheby},~\ref{eq:cheby_recursion}). Of course, the propagator is also limited to standard Hermitian dynamics. Lastly, as a baseline, we benchmark the direct optimization of the square-modulus functional for a $\sqrt{\text{\iSWAP}}$ gate with the analytic gradient in Eq.~\eqref{eq:grad_J_T_sm}, that is, without any use of automatic differentiation (``Direct (Cheby) (SM)'').

In the left column, panels~(a,~c,~e), we vary the number of levels at which we cut off the transmon qubit between $N_q = 3$ and $N_q=15$ levels, which we show in terms of the Hilbert space size, $N_H = N_q^2$ for two transmons. The gate duration is constant at 100~ns. In the right column, panels~(b, d, f), we vary the gate duration between 20~ns and 800~ns while keeping the number of transmon levels constant at 5 ($N_H = 25$). The step size of the time grid is 0.1~ns, so that the number of time steps in the scheme of Fig.~\ref{fig:schemes}~(a) is directly proportional to the gate duration and varies between 200 and 8000 steps.

For each optimization, we run the optimization for 10 iterations. At a minimum, each iteration includes one evaluation of the gradient as depicted in Fig.~\ref{fig:schemes}~(a), that is, a forward propagation of the four logical basis states and a backward propagation of four extended gradient states. However, the linesearch may require additional evaluations of the gradient in order to determine the step width. The number of linesearch steps varies for different optimizations and between iterations. Thus, the benchmark of the runtime in panels~(a,b) shows the seconds required per gradient evaluation. The optimizations were performed on an Intel Xeon Gold 6226R CPU workstation with a nominal clock speed of 2.9~GHz, without parallelization. The runtime was determined using \texttt{BenchmarkTools.jl}~\cite{BenchmarkTools-jl} as the minimum of 20 optimization runs, with a maximum total benchmark time of 24 hours. It excludes compilation overhead.

Fundamentally, the runtime is super-linear with the size of the Hilbert space. This is due to two factors: First, an increase in the spectral range, which for the Chebychev propagation implies an increase in the number of coefficients required for Eq.~\eqref{eq:cheby} to converge to machine precision. This increase is roughly linear with the number of levels in the transmon, i.e., the square root of the Hilbert space dimension. Second, the numerical scaling of the matrix-vector multiplications in Eq.~\eqref{eq:cheby_recursion} (and likewise in the implementation of a Runge-Kutta \acr{ODE} solver). In principle, matrix-vector multiplication scales as $N_H^2$, although this is mitigated by the fact that we use sparse matrices to store the Hamiltonian. For the gate duration, the runtime is directly proportional to the number of time steps.

We find that the runtime of the Semi-AD optimization is virtually indistinguishable from that of the direct optimization. This is in stark contrast to the full-AD optimization with a general \acr{ODE} solver. For a Hilbert space dimension greater than 64, the runtime of this optimization becomes prohibitively expensive (greater than 10 minutes per iteration, or step in the line search). The situation improves dramatically when using a modified Chebychev propagator for the full-AD optimization. Within the shown parameter region, these optimizations are well within the order of one minute per iteration. However, the insets in panels~(a, b) show a significant difference in scaling between the semi-AD and full-AD Chebychev optimizations. For the linear scaling in panel~(b), we find a factor of 10 between the different slopes (3~ms per time step versus 30~ms per time step). This order of magnitude also appears realistic for the scaling with respect to the Hilbert space dimension, and may become a problem for the runtime of much larger systems. Certainly, it will not be possible to perform a full-AD optimization of anything but the most trivial open quantum systems, with a Liouville space that is quadratically larger than the underlying Hilbert space. This is especially true because the Chebychev propagator is not applicable to non-unitary (open system) dynamics. The use of Newton polynomials discussed in Section~\ref{sec:grad_eval} is numerically more demanding than the use of Chebychev polynomials, and it is unclear whether it would be feasible to implement a variant of the method that avoids in-place operations and would be compatible with a framework for automatic differentiation such as Zygote. Thus, a full-AD optimization of an open quantum system would likely have to rely on an \acr{ODE} solver, with the prohibitive runtime shown in panel~(a).
The relative scaling of a direct optimization compared to a full-AD optimization matches recent observations in Ref.~\cite{LuAPS2022}.

There is no substantial difference in runtime between the different optimization functionals. This illustrates that the numerical effort of the optimization is entirely dominated by the time propagation, and further explains why the performance of the Semi-AD optimization is indistinguishable from a direct optimization with an analytic gradient.

The peak \acr{RAM} usage shown in panels~(c, d) was measured by monitoring the Julia process running the optimization with \texttt{psutil}~\cite{psutil}, and subtracting a baseline from a ``Hello World'' program to account for the footprint of the Julia runtime, which can vary depending on the exact version of Julia and the installed packages. We show the median \acr{RAM} usage, as well as the range of values in the inset for the Semi-AD (Cheby) optimization. There is significant fluctuation in the peak \acr{RAM} usage, as it depends on Julia's garbage collector, which has some element of randomness. However, the data shown in the insets indicates that the peak \acr{RAM} usage for semi-automatic differentiation is essentially constant around 100~MB, and equal to the memory used for the direct optimization with an analytic gradient. In principle, as analyzed in Section~\ref{sec:memory}, we would expect a linear scaling for large Hilbert space dimensions,. This is determined by the storage of the forward-propagated states marked in red in Fig.~\ref{fig:schemes}~(a). For $N=4$ targets, a Hilbert space dimension of $N_H$, and $N_T$ time steps, each stored state is a vector of complex numbers of length $N_H$, and each complex number requires 64~bits for both the real and imaginary part, i.e., 16~bytes in total. Thus, the total memory required for storage is
\begin{equation}
  \text{MB}_{\text{storage}} = N \cdot N_H \cdot N_T \cdot \frac{16}{(1024)^2}\,.
\end{equation}
For $N_H=225$, $N_T=1000$, the rightmost point in the inset of panel~(a), this comes out to 14~MB, and 49~MB for $N_H=100$, $N_T=8000$, the rightmost point in the inset of panel~(b), and we can conclude that for the data shown in Fig.~\ref{fig:benchmark}, the propagation overhead still dominates over the storage of propagated states.

The memory usage a full-AD optimization with an \acr{ODE} solver appears constant with respect to the size of the Hilbert space, albeit with a significant overhead, averaging around 700~MB. This is likely due to the solver being optimized for automatic differentiation, that is, providing handwritten adjoints for the Runge-Kutta step and thus eliminating the overhead of the computational graph for a single propagation step. Memory usage still scales linearly with the number of time steps, reaching 4~GB for 8000 time steps. The full-AD optimization using a Chebychev expansion does not benefit from the AD-aware propagation in \texttt{DifferentialEquations.jl}, and thus the memory usage scales with the number of coefficients in Eq.~\eqref{eq:cheby}, resulting in an excessive \acr{RAM} usage of 8~GB for a Hilbert space dimension of 225. Thus, the use of the Chebychev propagator for significantly larger Hilbert spaces would be prohibitive, and a possible implementation of a Newton propagator would likely perform even worse, making the use of full-AD for open quantum systems impractical.

Lastly, in panels~(e, f) we show the accumulated memory allocated on the heap, as measured by \texttt{BenchmarkTools.jl}. Note the scale of the y-axis, which reaches $4\times10^5$~MB, i.e., 400~GB. This is normalized by the number of gradient evaluations. The allocations differ from the peak \acr{RAM} usage due to Julia's garbage collector, but correlate strongly with the runtime shown in panels~(a, b). This illustrates the importance of good memory management, which is easy for an in-place Chebychev propagator (the negligible allocations of $<$~60~MB shown in the inset), but impossible for a propagator running in an automatic differentiation framework that does not allow for in-place operations.

\section{Conclusion and Outlook}%
\label{sec:conclusion}

We have developed a theory of semi-automatic differentiation that allows to optimize arbitrary functionals in an efficient and flexible manner.  Separating time propagation and the evaluation of the functional eliminates the excessive computational overhead and limited scope traditionally associated with the use of automatic differentiation (AD). With semi-AD, the time propagation and the associated gradient can be evaluated outside of the AD framework in a modified \acr{GRAPE} scheme, which we have described in detail. This scheme can be implemented in the most efficient manner possible, using sparse linear algebra, complex matrices, and in-place operations. The remaining part of the gradient is of minimal computational complexity and can thus be efficiently evaluated within an AD framework.

For the example of entangling gates on superconducting transmon qubits, we have verified that we can optimize for an arbitrary perfectly entangling quantum gate, either via a functional exploiting the geometric structure of the Weyl chamber, or by directly by maximizing the gate concurrence. This is the first demonstration of a gradient-based optimization of the concurrence, or any non-analytic functional. Fundamentally, such functionals \emph{require} the use of automatic differentiation and have been held back thus far by the associated numerical overhead.

The runtime and memory usage of the semi-AD optimizations shown here scales identically to a direct optimization of a quantum gate with a fully analytic gradient. That is, we completely eliminate the exorbitant numerical overhead traditionally associated with automatic differentiation. A ``full-AD'' optimization that uses a generic \acr{ODE} solver within the AD framework becomes unfeasible in terms of runtime for Hilbert space dimensions greater than $\approx 100$. The runtime can be improved significantly by adapting a propagation via expansion into Chebychev polynomials to the requirements of the AD framework (no in-place linear algebra operations). However, this results in excessive memory usage and would be difficult to extend to the significantly more complicated propagators required for open quantum systems.

As we have demonstrated, the potential improvements of semi-automatic differentiation scale super-linearly both for runtime and memory usage, compared to a full-AD approach. Even for moderate Hilbert space sizes in a closed quantum system, we observe improvements of up to two orders of magnitude. For control problems of larger dimension (either the dimension of the Hilbert space or the number of control parameters), and especially in open quantum systems, that effect will be further magnified. In such a setting, and for functionals where an analytic gradient is not feasible, semi-automatic differentiation will be the only viable option. Fundamentally, an optimization will be possible as long as simulating the time dynamics of the system is computationally feasible.

We have implemented the semi-AD approach in two ready-to-use Julia packages, \texttt{GRAPE.jl}~\cite{GRAPE-jl} and \texttt{Krotov.jl}~\cite{Krotov-jl}, as part of the more general \texttt{QuantumControl.jl}~\cite{QuantumControl-jl}. As an AD framework, we have used Zygote~\cite{Innes1810.07951, Zygote}. However, the method described here is applicable to any language or AD framework. In fact, the low complexity of the evaluation of the optimization functional relative to the full time propagation allows for additional avenues in environments where AD is underdeveloped. For example, we have tested the use of finite differences~\cite{FiniteDifferences-jl} and found it to be adequate for the examples in Section~\ref{sec:transmon}. Similarly, we would expect the easier-to-implement forward-mode differentiation to be a practical alternative.

We have developed the method of semi-automatic differentiation for the standard \acr{GRAPE} model of piecewise-constant control fields, albeit extending it to arbitrary functionals. However, the idea applies also to extensions of \acr{GRAPE} for a reduced number of control parameters, as used in the \acr{GOAT}~\cite{MachnesPRL2018}, \acr{GROUP}~\cite{SorensenPRA2018}, and \acr{GRAFS}~\cite{LucarelliPRA2018} methods. Parametrization of the control field may be taken into account by a further chain rule in Eq.~\eqref{eq:gradtau}. Furthermore, as pointed out in Ref.~\cite{MachnesPRL2018}, Eq.~\eqref{eq:gradprop} can be used not just to evaluate the gradient of a piecewise constant time step, but also the gradient a full time propagation with respect to a single control parameter. In all of these cases, the method can be augmented with semi-automatic differentiation to extend it to arbitrary functionals. We will explore this in future work as part of the \texttt{QuantumControl.jl} framework~\cite{QuantumControl-jl}.

In addition to enabling the use of AD, a second motivation for using a framework like Tensorflow to model the entire optimization problem was to enable the use of \acr{GPU} computing, enabling considerable speedups~\cite{LeungPRA2017,Ball2001.04060}. This possibility remains with semi-automatic differentiation, but the use of AD and \acr{GPU} computing are now entirely independent: The time propagation can be implemented in whatever way is most efficient, including on the \acr{GPU}.

Lastly, in Ref.~\cite{GoodwinJCP2016}, it was observed that the scheme in Fig.~\ref{fig:schemes}~(a) can be extended to compute the full Hessian of an overlap of two states. With next-generation AD systems that allow for the calculation of higher-order derivatives~\cite{Diffractor-jl}, this opens up the possibility of a full Hessian semi-automatic differentiation approach. The resulting Newton optimization may provide better convergence than the pseudo-Newton achievable via \acr{LBFGS} that we have used here.

Application of the semi-automatic differentiation framework developed in this paper will open new pathways to solve long-standing problems in quantum control. Apart from the optimization of entanglement measures in quantum information, which we have demonstrated here, and which could be extended to open quantum systems~\cite{ReichPhD2015}, the method would be applicable to the creation of multipartite entanglement in many-body systems. To date, this has been addressed with optimal control, either indirectly~\cite{PlatzerPRL2010} or with gradient-free methods~\cite{CanevaNJP2012}, usually via the \acr{CRAB} method~\cite{DoriaPRL2011,CanevaPRA2011,RachPRA2015}. The ability to explore the optimization landscape for these control problems more broadly may lead to deep insights into the quantum behavior, e.g., of biological systems~\cite{LeeS2007, HuelgaCP2013}.

In quantum metrology, there is a wide range of opportunities to use optimal control beyond simple state-to-state transfers. For one, we can address functionals like the recently developed population transfer functional for atom interferometry~\cite{GoerzSPIEO2021}. So far, this has only been explored with Krotov's method, but the use of semi-AD would allow optimizing the components of an atom interferometer using \acr{GRAPE}, potentially exploiting the
improved asymptotic convergence~\cite{MachnesPRA2011} in the generally flat optimization landscape of a robustness optimization~\cite{GoerzPRA2014}. This would also extend to a recently proposed  tractor atom interferometer~\cite{DuspayevPRA2021, RaithelQST2022}. More fundamentally, we may directly maximize metrological measures through quantum control~\cite{SekatskiQ2017,LinPRA2021}. Generalizing from a recent application of optimal control to the creation of extreme spin-squeezed states~\cite{CarrascoPRA2022} we may wish to directly maximize the quantum Fisher information~\cite{BraunsteinPRL1994,PezzePRL2009}, which is non-analytic in open quantum systems~\cite{MaPR2011}. Thus, we expect semi-automatic differentiation to be an indispensable tool for the design of metrological protocols in open quantum systems.

\section*{Data Availability}%

The code used to generate the benchmarks in Section~\ref{sec:transmon} and Fig.~\ref{fig:benchmark} is available at Ref.~\cite{semiad_data}, or under \href{https://doi.org/10.5281/zenodo.7386493}{DOI 10.5281/zenodo.7386493}. Moreover, examples for the perfect entangler and concurrence optimizations for coupled transmon qubits that show the resulting optimized fields are available as part of the documentation of the \texttt{GRAPE.jl} and \texttt{Krotov.jl} packages, which implement the semi-AD approach~\cite{GRAPE-jl, Krotov-jl}.

\section*{Acknowledgments}%

MHG and SCC acknowledge support by the DEVCOM Army Research Laboratory under
Cooperative Agreement Number W911NF-16-2-0147 and W911NF-21-2-0037, respectively. The work was also supported by DEVCOM Army Research Laboratory through DIRA-TRC No.\ DTR19-CI-019. The authors thank Alastair Marshall for contributions to the \texttt{GRAPE.jl} package.

\bibliographystyle{quantum}
\bibliography{refs}

\begin{thebibliography}{100}

\bibitem{BrifNJP2010}
Constantin Brif, Raj Chakrabarti, and Herschel Rabitz.
\newblock ``Control of quantum phenomena: past, present and future''.
\newblock \href{https://dx.doi.org/10.1088/1367-2630/12/7/075008}{New J. Phys.
  {\bf 12}, 075008}~(2010).

\bibitem{Shapiro2012}
Moshe Shapiro and Paul Brumer.
\newblock ``Quantum control of molecular processes''.
\newblock \href{https://dx.doi.org/10.1002/9783527639700}{Wiley \& Sons}.
  ~(2012).
\newblock Second edition.

\bibitem{KochJPCM2016}
Christiane~P. Koch.
\newblock ``Controlling open quantum systems: tools, achievements, and
  limitations''.
\newblock \href{https://dx.doi.org/10.1088/0953-8984/28/21/213001}{J. Phys.:
  Condens. Matter {\bf 28}, 213001}~(2016).

\bibitem{SolaAAMOP2018}
Ignacio~R. Sola, Bo~Y. Chang, Svetlana~A. Malinovskaya, and Vladimir~S.
  Malinovsky.
\newblock ``Quantum control in multilevel systems''.
\newblock In Ennio Arimondo, Louis~F. DiMauro, and Susanne~F. Yelin, editors,
  Advances In Atomic, Molecular, and Optical Physics.
\newblock \href{https://dx.doi.org/10.1016/bs.aamop.2018.02.003}{Volume~67,
  chapter~3, pages 151--256}.
\newblock Academic Press~(2018).

\bibitem{MorzhinRMS2019}
Oleg~V. Morzhin and Alexander~N. Pechen.
\newblock ``Krotov method for optimal control of closed quantum systems''.
\newblock \href{https://dx.doi.org/10.1070/rm9835}{Russ. Math. Surv. {\bf 74},
  851}~(2019).

\bibitem{Wilhelm2003.10132}
Frank~K. Wilhelm, Susanna Kirchhoff, Shai Machnes, Nicolas Wittler, and
  Dominique Sugny.
\newblock ``An introduction into optimal control for quantum
  technologies''~(2020).
\newblock  \href{http://arxiv.org/abs/2003.10132}{arXiv:2003.10132}.

\bibitem{KochEPJQT2022}
Christiane~P. Koch, Ugo Boscain, Tommaso Calarco, Gunther Dirr, Stefan Filipp,
  Steffen~J. Glaser, Ronnie Kosloff, Simone Montangero, Thomas
  Schulte-Herbrüggen, Dominique Sugny, and Frank~K. Wilhelm.
\newblock ``Quantum optimal control in quantum technologies. strategic report
  on current status, visions and goals for research in {Europe}''.
\newblock \href{https://dx.doi.org/10.1140/epjqt/s40507-022-00138-x}{EPJ
  Quantum Technol. {\bf 9}, 19}~(2022).

\bibitem{NielsenChuang2000}
Michael Nielsen and Isaac~L. Chuang.
\newblock ``Quantum computation and quantum information''.
\newblock \href{https://dx.doi.org/10.1017/CBO9780511976667}{Cambridge
  University Press}. ~(2000).

\bibitem{GeorgescuRMP2014}
Iulia~M. Georgescu, Sahel Ashhab, and Franco Nori.
\newblock ``Quantum simulation''.
\newblock \href{https://dx.doi.org/10.1103/RevModPhys.86.153}{Rev. Mod. Phys.
  {\bf 86}, 153}~(2014).

\bibitem{DegenRMP2017}
Christian~L. Degen, Friedemann Reinhard, and Paola Cappellaro.
\newblock ``Quantum sensing''.
\newblock \href{https://dx.doi.org/10.1103/RevModPhys.89.035002}{Rev. Mod.
  Phys. {\bf 89}, 035002}~(2017).

\bibitem{BoscainPRXQ2021}
Ugo Boscain, Mario Sigalotti, and Dominique Sugny.
\newblock ``Introduction to the {Pontryagin} maximum principle for quantum
  optimal control''.
\newblock \href{https://dx.doi.org/10.1103/prxquantum.2.030203}{PRX Quantum
  {\bf 2}, 030203}~(2021).

\bibitem{KhanejaJMR2005}
Navin Khaneja, Timo Reiss, Cindie Kehlet, Thomas Schulte-Herbrüggen, and
  Steffen~J. Glaser.
\newblock ``Optimal control of coupled spin dynamics: design of {NMR} pulse
  sequences by gradient ascent algorithms''.
\newblock \href{https://dx.doi.org/10.1016/j.jmr.2004.11.004}{J. Magnet. Res.
  {\bf 172}, 296}~(2005).

\bibitem{FouquieresJMR2011}
Pierre de~Fouquières, Sophie~G. Schirmer, Steffen~J. Glaser, and Ilya Kuprov.
\newblock ``Second order gradient ascent pulse engineering''.
\newblock \href{https://dx.doi.org/10.1016/j.jmr.2011.07.023}{J. Magnet. Res.
  {\bf 212}, 412}~(2011).

\bibitem{KrotovEC1983}
Vadim~F. Krotov and N.~N. Fel'dman.
\newblock ``An iterative method for solving optimal-control problems''.
\newblock Engrg. Cybernetics {\bf 21}, 123~(1983).

\bibitem{KrotovCC1988}
Vadim~F. Krotov.
\newblock ``A technique of global bounds in optimal control theory''.
\newblock Control Cybern. {\bf 17}, 115~(1988).

\bibitem{KrotovBook}
Vadim~F. Krotov.
\newblock ``Global methods in optimal control''.
\newblock Dekker. New York, NY, USA~(1996).

\bibitem{SomloiCP1993}
József Somlói, Vladimir~A. Kazakov, and David~J. Tannor.
\newblock ``Controlled dissociation of {I$_2$} via optical transitions between
  the {X} and {B} electronic states''.
\newblock \href{https://dx.doi.org/10.1016/0301-0104(93)80108-L}{Chem. Phys.
  {\bf 172}, 85}~(1993).

\bibitem{BartanaJCP1997}
Allon Bartana, Ronnie Kosloff, and David~J. Tannor.
\newblock ``Laser cooling of internal degrees of freedom. {II}''.
\newblock \href{https://dx.doi.org/10.1063/1.473973}{J. Chem. Phys. {\bf 106},
  1435}~(1997).

\bibitem{PalaoPRA2003}
José~P. Palao and Ronnie Kosloff.
\newblock ``Optimal control theory for unitary transformations''.
\newblock \href{https://dx.doi.org/10.1103/PhysRevA.68.062308}{Phys. Rev. A
  {\bf 68}, 062308}~(2003).

\bibitem{ReichJCP2012}
Daniel~M. Reich, Mamadou Ndong, and Christiane~P. Koch.
\newblock ``Monotonically convergent optimization in quantum control using
  {Krotov's} method''.
\newblock \href{https://dx.doi.org/10.1063/1.3691827}{J. Chem. Phys. {\bf 136},
  104103}~(2012).

\bibitem{RashidinejadIJQE2016}
Amir Rashidinejad, Yihan Li, and Andrew~M. Weiner.
\newblock ``Recent advances in programmable photonic-assisted ultrabroadband
  radio-frequency arbitrary waveform generation''.
\newblock \href{https://dx.doi.org/10.1109/jqe.2015.2506987}{IEEE J. Quantum
  Electron. {\bf 52}, 1}~(2016).

\bibitem{WeinerRSI2000}
Andrew~M. Weiner.
\newblock ``Femtosecond pulse shaping using spatial light modulators''.
\newblock \href{https://dx.doi.org/10.1063/1.1150614}{Rev. Sci. Instr. {\bf
  71}, 1929}~(2000).

\bibitem{JohanssonCPC2013}
J.~Robert Johansson, Paul~D. Nation, and Franco Nori.
\newblock ``{QuTiP 2}: A {Python} framework for the dynamics of open quantum
  systems''.
\newblock \href{https://dx.doi.org/10.1016/j.cpc.2012.11.019}{Comput. Phys.
  Commun. {\bf 184}, 1234}~(2013).

\bibitem{MachnesPRA2011}
Shai Machnes, Uwe Sander, Steffen~J. Glaser, Pierre de~Fouquières, Audrūnas
  Gruslys, Sophie~G. Schirmer, and Thomas Schulte-Herbrüggen.
\newblock ``Comparing, optimizing, and benchmarking quantum-control algorithms
  in a unifying programming framework''.
\newblock \href{https://dx.doi.org/10.1103/PhysRevA.84.022305}{Phys. Rev. A
  {\bf 84}, 022305}~(2011).

\bibitem{HogbenJMR2011}
Hannah~J. Hogben, Matthew Krzystyniak, Gareth T.~P. Charnock, Peter~J. Hore,
  and Ilya Kuprov.
\newblock ``Spinach -- a software library for simulation of spin dynamics in
  large spin systems''.
\newblock \href{https://dx.doi.org/10.1016/j.jmr.2010.11.008}{J. Magnet. Res.
  {\bf 208}, 179}~(2011).

\bibitem{TosnerJMR2009}
Zdeněk Tošner, Thomas Vosegaard, Cindie Kehlet, Navin Khaneja, Steffen~J.
  Glaser, and Niels~Chr. Nielsen.
\newblock ``Optimal control in {NMR} spectroscopy: Numerical implementation in
  {SIMPSON}''.
\newblock
  \href{https://dx.doi.org/http://dx.doi.org/10.1016/j.jmr.2008.11.020}{J.
  Magnet. Res. {\bf 197}, 120}~(2009).

\bibitem{GoerzSPP2019}
Michael~H. Goerz, Daniel Basilewitsch, Fernando Gago-Encinas, Matthias~G.
  Krauss, Karl~P. Horn, Daniel~M. Reich, and Christiane~P. Koch.
\newblock ``Krotov: A {Python} implementation of {Krotov}'s method for quantum
  optimal control''.
\newblock \href{https://dx.doi.org/10.21468/scipostphys.7.6.080}{SciPost Phys.
  {\bf 7}, 080}~(2019).

\bibitem{BraunsteinPRL1994}
Samuel~L. Braunstein and Carlton~M. Caves.
\newblock ``Statistical distance and the geometry of quantum states''.
\newblock \href{https://dx.doi.org/10.1103/physrevlett.72.3439}{Phys. Rev.
  Lett. {\bf 72}, 3439}~(1994).

\bibitem{PezzePRL2009}
Luca Pezzé and Augusto Smerzi.
\newblock ``Entanglement, nonlinear dynamics, and the {Heisenberg} limit''.
\newblock \href{https://dx.doi.org/10.1103/physrevlett.102.100401}{Phys. Rev.
  Lett. {\bf 102}, 100401}~(2009).

\bibitem{MaPR2011}
Jian Ma, Xiaoguang Wang, Changpu~P. Sun, and Franco Nori.
\newblock ``Quantum spin squeezing''.
\newblock \href{https://dx.doi.org/10.1016/j.physrep.2011.08.003}{Phys. Rep.
  {\bf 509}, 89}~(2011).

\bibitem{KrausPRA2001}
Barbara Kraus and J.~Ignacio Cirac.
\newblock ``Optimal creation of entanglement using a two-qubit gate''.
\newblock \href{https://dx.doi.org/10.1103/PhysRevA.63.062309}{Phys. Rev. A
  {\bf 63}, 062309}~(2001).

\bibitem{ZhangPRA2003}
Jun Zhang, Jiří Vala, Shankar Sastry, and K.~Birgitta Whaley.
\newblock ``Geometric theory of nonlocal two-qubit operations''.
\newblock \href{https://dx.doi.org/10.1103/PhysRevA.67.042313}{Phys. Rev. A
  {\bf 67}, 042313}~(2003).

\bibitem{WattsE2013}
Paul Watts, Maurice O'Connor, and Jiří Vala.
\newblock ``Metric structure of the space of two-qubit gates, perfect
  entanglers and quantum control''.
\newblock \href{https://dx.doi.org/10.3390/e15061963}{Entropy {\bf 15},
  1963}~(2013).

\bibitem{WattsPRA2015}
Paul Watts, Jiří Vala, Matthias~M. Müller, Tommaso Calarco, K.~Birgitta
  Whaley, Daniel~M. Reich, Michael~H. Goerz, and Christiane~P. Koch.
\newblock ``Optimizing for an arbitrary perfect entangler: {I. Functionals}''.
\newblock \href{https://dx.doi.org/10.1103/PhysRevA.91.062306}{Phys. Rev. A
  {\bf 91}, 062306}~(2015).

\bibitem{GoerzPRA2015}
Michael~H. Goerz, Giulia Gualdi, Daniel~M. Reich, Christiane~P. Koch, Felix
  Motzoi, K.~Birgitta Whaley, Jiří Vala, Matthias~M. Müller, Simone
  Montangero, and Tommaso Calarco.
\newblock ``Optimizing for an arbitrary perfect entangler. {II. Application}''.
\newblock \href{https://dx.doi.org/10.1103/PhysRevA.91.062307}{Phys. Rev. A
  {\bf 91}, 062307}~(2015).

\bibitem{JirariEL2009}
Hamza Jirari.
\newblock ``Optimal control approach to dynamical suppression of decoherence of
  a qubit''.
\newblock \href{https://dx.doi.org/10.1209/0295-5075/87/40003}{Europhys. Lett.
  {\bf 87}, 40003}~(2009).

\bibitem{LeungPRA2017}
Nelson Leung, Mohamed Abdelhafez, Jens Koch, and David~I. Schuster.
\newblock ``Speedup for quantum optimal control from automatic differentiation
  based on graphics processing units''.
\newblock \href{https://dx.doi.org/10.1103/PhysRevA.95.042318}{Phys. Rev. A
  {\bf 95}, 042318}~(2017).

\bibitem{AbdelhafezPRA2019}
Mohamed Abdelhafez, David~I. Schuster, and Jens Koch.
\newblock ``Gradient-based optimal control of open quantum systems using
  quantum trajectories and automatic differentiation''.
\newblock \href{https://dx.doi.org/10.1103/PhysRevA.99.052327}{Phys. Rev. A
  {\bf 99}, 052327}~(2019).

\bibitem{AbdelhafezPRA2020}
Mohamed Abdelhafez, Brian Baker, András Gyenis, Pranav Mundada, Andrew~A.
  Houck, David~I. Schuster, and Jens Koch.
\newblock ``Universal gates for protected superconducting qubits using optimal
  control''.
\newblock \href{https://dx.doi.org/10.1103/physreva.101.022321}{Phys. Rev. A
  {\bf 101}, 022321}~(2020).

\bibitem{SchaeferMLST2020}
Frank Schäfer, Michal Kloc, Christoph Bruder, and Niels Lörch.
\newblock ``A differentiable programming method for quantum control''.
\newblock \href{https://dx.doi.org/10.1088/2632-2153/ab9802}{Mach. Learn.: Sci.
  Technol. {\bf 1}, 035009}~(2020).

\bibitem{quantum-optimal-control}
Nelson Leung and {Contributors}~(2021).
\newblock
  code:~\href{https://github.com/SchusterLab/quantum-optimal-control}{SchusterLab/quantum-optimal-control}.

\bibitem{qoc}
Daniel Weiss and {Contributors}~(2021).
\newblock  code:~\href{https://github.com/SchusterLab/qoc}{SchusterLab/qoc}.

\bibitem{Griewank2008}
Andreas Griewank and Andrea Walther.
\newblock ``Evaluating derivatives''.
\newblock \href{https://dx.doi.org/10.1137/1.9780898717761}{Society for
  Industrial and Applied Mathematics}. Philadelphia~(2008).
\newblock Second edition.

\bibitem{MargossianWIDM2019}
Charles~C. Margossian.
\newblock ``A review of automatic differentiation and its efficient
  implementation''.
\newblock \href{https://dx.doi.org/10.1002/widm.1305}{WIREs Data Mining Knowl
  Discov.{\bf 9}}~(2019).

\bibitem{SchaeferMLST2021}
Frank Schäfer, Pavel Sekatski, Martin Koppenhöfer, Christoph Bruder, and
  Michal Kloc.
\newblock ``Control of stochastic quantum dynamics by differentiable
  programming''.
\newblock \href{https://dx.doi.org/10.1088/2632-2153/abec22}{Mach. Learn.: Sci.
  Technol. {\bf 2}, 035004}~(2021).

\bibitem{PropsonPRA2022}
Thomas Propson, Brian~E. Jackson, Jens Koch, Zachary Manchester, and David~I.
  Schuster.
\newblock ``Robust quantum optimal control with trajectory optimization''.
\newblock \href{https://dx.doi.org/10.1103/physrevapplied.17.014036}{Phys. Rev.
  Applied {\bf 17}, 014036}~(2022).

\bibitem{GoerzQST2018}
Michael~H. Goerz and Kurt Jacobs.
\newblock ``Efficient optimization of state preparation in quantum networks
  using quantum trajectories''.
\newblock \href{https://dx.doi.org/10.1088/2058-9565/aace16}{Quantum Sci.
  Technol. {\bf 3}, 045005}~(2018).

\bibitem{WittlerPRA2021}
Nicolas Wittler, Federico Roy, Kevin Pack, Max Werninghaus, Anurag~Saha Roy,
  Daniel~J. Egger, Stefan Filipp, Frank~K. Wilhelm, and Shai Machnes.
\newblock ``Integrated tool set for control, calibration, and characterization
  of quantum devices applied to superconducting qubits''.
\newblock \href{https://dx.doi.org/10.1103/physrevapplied.15.034080}{Phys. Rev.
  Applied {\bf 15}, 034080}~(2021).

\bibitem{Ball2001.04060}
Harrison Ball, Michael~J. Biercuk, Andre Carvalho, Jiayin Chen, Michael Hush,
  Leonardo A.~De Castro, Li~Li, Per~J. Liebermann, Harry~J. Slatyer, Claire
  Edmunds, Virginia Frey, Cornelius Hempel, and Alistair Milne.
\newblock ``Software tools for quantum control: Improving quantum computer
  performance through noise and error suppression''~(2020).
\newblock  \href{http://arxiv.org/abs/2001.04060}{arXiv:2001.04060}.

\bibitem{qgrad}
Asad Raza and {Contributors}~(2022).
\newblock  code:~\href{https://github.com/qgrad/qgrad}{qgrad/qgrad}.

\bibitem{GriewankISMP2012}
Andreas Griewank.
\newblock ``Who invented the reverse mode of differentiation''.
\newblock In Martin Grötschel, editor, Optimization Stories.
\newblock Pages 389--400.
\newblock Documenta Mathematica. 21st International Symposium on Mathematical
  Programming, Berlin~(2012).
\newblock
  url:~\href{https://www.zib.de/groetschel/publications/OptimizationStories.pdf}{www.zib.de/groetschel/publications/OptimizationStories.pdf}.

\bibitem{RumelhartN1986}
David~E. Rumelhart, Geoffrey~E. Hinton, and Ronald~J. Williams.
\newblock ``Learning representations by back-propagating errors''.
\newblock \href{https://dx.doi.org/10.1038/323533a0}{Nature {\bf 323},
  533}~(1986).

\bibitem{Tensorflow}
Martin Abadi, Paul Barham, Jianmin Chen, Zhifeng Chen, Andy Davis, Jeffrey
  Dean, Matthieu Devin, Sanjay Ghemawat, Geoffrey Irving, Michael Isard,
  Manjunath Kudlur, Josh Levenberg, Rajat Monga, Sherry Moore, Derek~G. Murray,
  Benoit Steiner, Paul Tucker, Vijay Vasudevan, Pete Warden, Martin Wicke, Yuan
  Yu, and Xiaoqiang Zheng.
\newblock ``Tensorflow: A system for large-scale machine learning''.
\newblock In 12th USENIX Symposium on Operating Systems Design and
  Implementation (OSDI 16).
\newblock Page 265.
\newblock ~(2016).
\newblock  url:~\href{https://www.tensorflow.org/}{www.tensorflow.org/}.

\bibitem{PaszkeNIPS2019}
Adam Paszke, Sam Gross, Francisco Massa, Adam Lerer, James Bradbury, Gregory
  Chanan, Trevor Killeen, Zeming Lin, Natalia Gimelshein, Luca Antiga, Alban
  Desmaison, Andreas Köpf, Edward Yang, Zachary DeVito, Martin Raison, Alykhan
  Tejani, Sasank Chilamkurthy, Benoit Steiner, Lu~Fang, Junjie Bai, and Soumith
  Chintala.
\newblock ``{PyTorch}: An imperative style, high-performance deep learning
  library''.
\newblock In Hanna~M. Wallach, Hugo Larochelle, Alina Beygelzimer, Florence
  d'Alché Buc, Edward~A. Fox, and Roman Garnett, editors, Advances in Neural
  Information Processing Systems 32.
\newblock Pages 8024--8035.
\newblock Vancouver, BC, Canada~(2019). Annual Conference on Neural Information
  Processing Systems 2019, NeurIPS 2019.
\newblock
  url:~\href{http://papers.neurips.cc/paper/9015-pytorch-an-imperative-style-high-performance-deep-learning-library.pdf}{http://papers.neurips.cc/paper/9015-pytorch-an-imperative-style-high-performance-deep-learning-library.pdf}.

\bibitem{pytorch}
Adam Paszke and {Contributors}~(2022).
\newblock  code:~\href{https://github.com/pytorch/pytorch}{pytorch/pytorch}.

\bibitem{FrostigSYSML2018}
Roy Frostig, Matthew Johnson, and Chris Leary.
\newblock ``Compiling machine learning programs via high-level tracing''.
\newblock In SysML Conference.
\newblock Stanford, CA~(2018).
\newblock
  url:~\href{https://mlsys.org/Conferences/2019/doc/2018/146.pdf}{mlsys.org/Conferences/2019/doc/2018/146.pdf}.

\bibitem{jax}
James Bradbury, Roy Frostig, Peter Hawkins, Matthew~James Johnson, Chris Leary,
  Dougal Maclaurin, George Necula, Adam Paszke, Jake Vander{P}las, Skye
  Wanderman-{M}ilne, and Qiao Zhang~(2018).
\newblock  code:~\href{https://github.com/google/jax}{google/jax}.

\bibitem{Innes1811.01457}
Michael Innes, Elliot Saba, Keno Fischer, Dhairya Gandhi, Marco~Concetto
  Rudilosso, Neethu~Mariya Joy, Tejan Karmali, Avik Pal, and Viral Shah.
\newblock ``Fashionable modelling with {Flux}''~(2018).
\newblock  \href{http://arxiv.org/abs/1811.01457}{arXiv:1811.01457}.

\bibitem{InnesJOSS2018}
Michael Innes.
\newblock ``Flux: Elegant machine learning with {Julia}''.
\newblock \href{https://dx.doi.org/10.21105/joss.00602}{J. Open Source Softw.
  {\bf 3}, 602}~(2018).

\bibitem{Innes1810.07951}
Michael Innes.
\newblock ``Don't unroll adjoint: Differentiating {SSA}-form programs''~(2018).
\newblock  \href{http://arxiv.org/abs/1810.07951}{arXiv:1810.07951}.

\bibitem{Zygote}
Michael Innes and {Contributors}~(2022).
\newblock  code:~\href{https://github.com/FluxML/Zygote.jl}{FluxML/Zygote.jl}.

\bibitem{Tal-EzerJCP1984}
Hillel Tal-Ezer and Ronnie Kosloff.
\newblock ``An accurate and efficient scheme for propagating the time dependent
  {Schrödinger} equation''.
\newblock \href{https://dx.doi.org/10.1063/1.448136}{J. Chem. Phys. {\bf 81},
  3967}~(1984).

\bibitem{KosloffJCP1988}
Ronnie Kosloff.
\newblock ``Time-dependent quantum-mechanical methods for molecular dynamics''.
\newblock \href{https://dx.doi.org/10.1021/j100319a003}{J. Chem. Phys. {\bf
  92}, 2087}~(1988).

\bibitem{BermanJPA1992}
Michael Berman, Ronnie Kosloff, and Hillel Tal-Ezer.
\newblock ``Solution of the time-dependent liouville-von neumann equation:
  dissipative evolution''.
\newblock \href{https://dx.doi.org/10.1088/0305-4470/25/5/031}{J. Phys. A {\bf
  25}, 1283}~(1992).

\bibitem{KosloffARPC94}
Ronnie Kosloff.
\newblock ``Propagation methods for quantum molecular dynamics''.
\newblock \href{https://dx.doi.org/10.1146/annurev.pc.45.100194.001045}{Annu.
  Rev. Phys. Chem. {\bf 45}, 145}~(1994).

\bibitem{AshkenaziJCP1995}
Guy Ashkenazi, Ronnie Kosloff, Sanford Ruhman, and Hillel Tal-Ezer.
\newblock ``Newtonian propagation methods applied to the photodissociation
  dynamics of {I}$_3^{-}$''.
\newblock \href{https://dx.doi.org/10.1063/1.469904}{J. Chem. Phys. {\bf 103},
  10005--10014}~(1995).

\bibitem{BezansonSIREV2017}
Jeff Bezanson, Alan Edelman, Stefan Karpinski, and Viral~B. Shah.
\newblock ``Julia: A fresh approach to numerical computing''.
\newblock \href{https://dx.doi.org/10.1137/141000671}{SIAM Rev. {\bf 59},
  65}~(2017).

\bibitem{Julia}
``The {Julia} programming language''.
\newblock  url:~\href{https://julialang.org}{julialang.org}.

\bibitem{GRAPE-jl}
Michael~H. Goerz and {Contributors}~(2022).
\newblock
  code:~\href{https://github.com/JuliaQuantumControl/GRAPE.jl}{JuliaQuantumControl/GRAPE.jl}.

\bibitem{Krotov-jl}
Michael~H. Goerz and {Contributors}~(2022).
\newblock
  code:~\href{https://github.com/JuliaQuantumControl/Krotov.jl}{JuliaQuantumControl/Krotov.jl}.

\bibitem{QuantumControl-jl}
Michael~H. Goerz and {Contributors}~(2022).
\newblock
  code:~\href{https://github.com/JuliaQuantumControl/QuantumControl.jl}{JuliaQuantumControl/QuantumControl.jl}.

\bibitem{JKochPRA2007}
Jens Koch, Terri~M. Yu, Jay Gambetta, Andrew~A. Houck, David~I. Schuster,
  Johannes Majer, Alexandre Blais, Michel~H. Devoret, Steven~M. Girvin, and
  Robert~J. Schoelkopf.
\newblock ``Charge-insensitive qubit design derived from the {Cooper} pair
  box''.
\newblock \href{https://dx.doi.org/10.1103/PhysRevA.76.042319}{Phys. Rev. A
  {\bf 76}, 042319}~(2007).

\bibitem{BlaisPRA2007}
Alexandre Blais, Jay Gambetta, A.~Wallraff, D.~I. Schuster, Steven~M. Girvin,
  M.~H. Devoret, and Robert~J. Schoelkopf.
\newblock ``Quantum-information processing with circuit quantum
  electrodynamics''.
\newblock \href{https://dx.doi.org/10.1103/PhysRevA.75.032329}{Phys. Rev. A
  {\bf 75}, 032329}~(2007).

\bibitem{GoerzNJP2014arxiv2}
Michael~H. Goerz, Daniel~M. Reich, and Christiane~P. Koch.
\newblock ``Optimal control theory for a unitary operation under dissipative
  evolution''~(2021).
\newblock  \href{http://arxiv.org/abs/1312.0111v2}{arXiv:1312.0111v2}.

\bibitem{MatlabOTB}
The MathWorks.
\newblock Natick, MA, USA.
\newblock ``Matlab optimization toolbox''.
\newblock ~(2018).

\bibitem{VirtanenNM2020}
Pauli Virtanen, Ralf Gommers, Travis~E. Oliphant, Matt Haberland, Tyler Reddy,
  David Cournapeau, Evgeni Burovski, Pearu Peterson, Warren Weckesser, Jonathan
  Bright, Stéfan~J. {van der Walt}, Matthew Brett, Joshua Wilson, K.~Jarrod
  Millman, Nikolay Mayorov, Andrew R.~J. Nelson, Eric Jones, Robert Kern, Eric
  Larson, CJ~Carey, İlhan Polat, Yu~Feng, Eric~W. Moore, Jake {VanderPlas},
  Denis Laxalde, Josef Perktold, Robert Cimrman, Ian Henriksen, E.~A. Quintero,
  Charles~R. Harris, Anne~M. Archibald, Antônio~H. Ribeiro, Fabian Pedregosa,
  Paul {van Mulbregt}, and {SciPy 1.0 Contributors}.
\newblock ``{SciPy} 1.0: Fundamental algorithms for scientific computing in
  {Python}''.
\newblock \href{https://dx.doi.org/10.1038/s41592-019-0686-2}{Nat. Methods {\bf
  17}, 261}~(2020).

\bibitem{SciPy}
Eric Jones, Travis Oliphant, Pearu Peterson, et~al.
\newblock ``{SciPy}: Open source scientific tools for {Python}''.
\newblock ~(2001--).
\newblock
  url:~\href{https://docs.scipy.org/doc/scipy/}{docs.scipy.org/doc/scipy/}.

\bibitem{MogensenJOSS2018}
Patrick~K. Mogensen and Asbjørn~N. Riseth.
\newblock ``{Optim}: A mathematical optimization package for {Julia}''.
\newblock \href{https://dx.doi.org/10.21105/joss.00615}{J. Open Source Softw.
  {\bf 3}, 615}~(2018).

\bibitem{ByrdSJSC1995}
Richard~H. Byrd, Peihuang Lu, Jorge Nocedal, and Ciyou Zhu.
\newblock ``A limited memory algorithm for bound constrained optimization''.
\newblock \href{https://dx.doi.org/10.1137/0916069}{SIAM J. Sci. Comput. {\bf
  16}, 1190}~(1995).

\bibitem{ZhuATMS97}
Ciyou Zhu, Richard~H. Byrd, Peihuang Lu, and Jorge Nocedal.
\newblock ``Algorithm 778: {L-BFGS-B}: {Fortran} subroutines for large-scale
  bound-constrained optimization''.
\newblock \href{https://dx.doi.org/10.1145/279232.279236}{ACM Trans. Math.
  Softw. {\bf 23}, 550}~(1997).

\bibitem{LBFGSB-jl}
Yupei Qi and {Contributors}~(2022).
\newblock  code:~\href{https://github.com/Gnimuc/LBFGSB.jl}{Gnimuc/LBFGSB.jl}.

\bibitem{Linesearches-jl}
Patrick~K. Mogensen, Asbjørn~N. Riseth, and {Contributors}~(2020).
\newblock
  code:~\href{https://github.com/JuliaNLSolvers/LineSearches.jl}{JuliaNLSolvers/LineSearches.jl}.

\bibitem{VanLoanITAC1978}
Charles~F. Van~Loan.
\newblock ``Computing integrals involving the matrix exponential''.
\newblock \href{https://dx.doi.org/10.1109/tac.1978.1101743}{IEEE Trans.
  Automat. Contr. {\bf 23}, 395}~(1978).

\bibitem{GoodwinJCP2015}
David~L. Goodwin and Ilya Kuprov.
\newblock ``Auxiliary matrix formalism for interaction representation
  transformations, optimal control, and spin relaxation theories''.
\newblock \href{https://dx.doi.org/10.1063/1.4928978}{J. Chem. Phys. {\bf 143},
  084113}~(2015).

\bibitem{GilSeguraTemme2007}
Amparo Gil, Javier Segura, and Nico~M. Temme.
\newblock ``Numerical methods for special functions''.
\newblock \href{https://dx.doi.org/10.1137/1.9780898717822}{Society for
  Industrial and Applied Mathematics}. ~(2007).

\bibitem{Tal-EzerSJSC2007}
Hillel Tal-Ezer.
\newblock ``On restart and error estimation for {Krylov} approximation of
  $w=f(a)v$''.
\newblock \href{https://dx.doi.org/10.1137/040617868}{SIAM J. Sci. Comput. {\bf
  29}, 2426}~(2007).

\bibitem{QuantumPropagators-jl}
Michael~H. Goerz and {Contributors}~(2022).
\newblock
  code:~\href{https://github.com/JuliaQuantumControl/QuantumPropagators.jl}{JuliaQuantumControl/QuantumPropagators.jl}.

\bibitem{GoerzNJP2014}
Michael~H. Goerz, Daniel~M. Reich, and Christiane~P. Koch.
\newblock ``Optimal control theory for a unitary operation under dissipative
  evolution''.
\newblock \href{https://dx.doi.org/10.1088/1367-2630/16/5/055012}{New J. Phys.
  {\bf 16}, 055012}~(2014).

\bibitem{GoerzPRA2014}
Michael~H. Goerz, Eli~J. Halperin, Jon~M. Aytac, Christiane~P. Koch, and
  K.~Birgitta Whaley.
\newblock ``Robustness of high-fidelity {Rydberg} gates with single-site
  addressability''.
\newblock \href{https://dx.doi.org/10.1103/PhysRevA.90.032329}{Phys. Rev. A
  {\bf 90}, 032329}~(2014).

\bibitem{NdongJCP2009}
Mamadou Ndong, Hillel Tal-Ezer, Ronnie Kosloff, and Christiane~P. Koch.
\newblock ``A {Chebychev} propagator for inhomogeneous {Schrödinger}
  equations''.
\newblock \href{https://dx.doi.org/10.1063/1.3098940}{J. Chem. Phys. {\bf 130},
  124108}~(2009).

\bibitem{GoerzPhd2015}
Michael~H. Goerz.
\newblock ``Optimizing robust quantum gates in open quantum systems''.
\newblock PhD thesis.
\newblock Universität Kassel.
\newblock ~(2015).
\newblock
  url:~\href{https://d-nb.info/1072259729/34}{d-nb.info/1072259729/34}.

\bibitem{RackauckasWeb2021.12.25}
Christopher Rackauckas.
\newblock ``Engineering trade-offs in automatic differentiation: from
  {TensorFlow} and {PyTorch} to {Jax} and {Julia}''.
\newblock
  url:~\href{http://www.stochasticlifestyle.com/engineering-trade-offs-in-automatic-differentiation-from-tensorflow-and-pytorch-to-jax-and-julia/}{http://www.stochasticlifestyle.com/engineering-trade-offs-in-automatic-differentiation-from-tensorflow-and-pytorch-to-jax-and-julia/}.

\bibitem{MagnusNeudecker2019}
Jan~R. Magnus and Heinz Neudecker.
\newblock ``Matrix differential calculus with applications in statistics and
  econometrics''.
\newblock \href{https://dx.doi.org/10.1002/9781119541219}{Wiley Series in
  Probability and Statistics}. Wiley. ~(2019).
\newblock Third edition.

\bibitem{Petersen2008}
Kaare~B. Petersen and Michael~S. Pedersen.
\newblock ``The matrix cookbook''.
\newblock Technical report.
\newblock Technical University of Denmark~(2012).
\newblock
  url:~\href{http://www2.imm.dtu.dk/pubdb/p.php?3274}{http://www2.imm.dtu.dk/pubdb/p.php?3274}.

\bibitem{Giles2008}
Mike~B. Giles.
\newblock ``Collected matrix derivative results for forward and reverse mode
  algorithmic differentiation''.
\newblock In Advances in Automatic Differentiation.
\newblock \href{https://dx.doi.org/10.1007/978-3-540-68942-3_4}{Volume~64,
  pages 35--44}.
\newblock Springer~(2008).

\bibitem{Giles2008b}
Mike~B. Giles.
\newblock ``An extended collection of matrix derivative results for forward and
  reverse mode automatic differentiation''.
\newblock Technical Report NA-08-01.
\newblock Oxford University Computing Laboratory~(2008).
\newblock
  url:~\href{https://people.maths.ox.ac.uk/gilesm/files/NA-08-01.pdf}{people.maths.ox.ac.uk/gilesm/files/NA-08-01.pdf}.

\bibitem{Hjorungnes2011}
Are Hjørungnes.
\newblock ``Complex-valued matrix derivatives: With applications in signal
  processing and communications''.
\newblock \href{https://dx.doi.org/10.1017/CBO9780511921490}{Cambridge
  University Press}. ~(2011).

\bibitem{BlackfordTMS2002}
L.~Susan Blackford, James Demmel, Jack~J. Dongarra, Ian~S. Duff, Sven
  Hammarling, Greg Henry, Michael Heroux, Linda Kaufman, Andrew Lumsdain,
  Antoine Petitet, Roldan Pozo, Karin Remington, and R.~Clint Whaley.
\newblock ``An updated set of basic linear algebra subprograms ({BLAS})''.
\newblock \href{https://dx.doi.org/10.1145/567806.567807}{ACM Trans. Math.
  Softw. {\bf 28}, 135}~(2002).

\bibitem{SchirmerNJP2011}
Sophie~G. Schirmer and Pierre de~Fouquières.
\newblock ``Efficient algorithms for optimal control of quantum dynamics: the
  {Krotov} method unencumbered''.
\newblock \href{https://dx.doi.org/10.1088/1367-2630/13/7/073029}{New J. Phys.
  {\bf 13}, 073029}~(2011).

\bibitem{PalaoPRA2008}
José~P. Palao, Ronnie Kosloff, and Christiane~P. Koch.
\newblock ``Protecting coherence in optimal control theory: State-dependent
  constraint approach''.
\newblock \href{https://dx.doi.org/10.1103/PhysRevA.77.063412}{Phys. Rev. A
  {\bf 77}, 063412}~(2008).

\bibitem{Narayanan2022}
Sri H.~K. Narayanan, Thomas Propson, Marcelo Bongarti, Jan Hueckelheim, and
  Paul Hovland.
\newblock ``Reducing memory requirements of quantum optimal control''.
\newblock \href{https://dx.doi.org/10.48550/arXiv.2203.12717}{Technical Report
  ANL/MCS-P9566-0222}.
\newblock Argonne National Laboratory~(2022).

\bibitem{PolettoPRL2012}
Stefano Poletto, Jay~M. Gambetta, Seth~T. Merkel, John~A. Smolin, Jerry~M.
  Chow, A.~D. C\'orcoles, George~A. Keefe, Mary~B. Rothwell, J.~R. Rozen, D.~W.
  Abraham, Chad Rigetti, and M.~Steffen.
\newblock ``Entanglement of two superconducting qubits in a waveguide cavity
  via monochromatic two-photon excitation''.
\newblock \href{https://dx.doi.org/10.1103/PhysRevLett.109.240505}{Phys. Rev.
  Lett. {\bf 109}, 240505}~(2012).

\bibitem{GoerzNPJQI2017}
Michael~H. Goerz, Felix Motzoi, K.~Birgitta Whaley, and Christiane~P. Koch.
\newblock ``Charting the circuit {QED} design landscape using optimal control
  theory''.
\newblock \href{https://dx.doi.org/10.1038/s41534-017-0036-0}{npj Quantum Inf
  {\bf 3}, 37}~(2017).

\bibitem{MaSB2021}
Wen-Long Ma, Shruti Puri, Robert~J. Schoelkopf, Michel~H. Devoret, Steven~M.
  Girvin, and Liang Jiang.
\newblock ``Quantum control of bosonic modes with superconducting circuits''.
\newblock \href{https://dx.doi.org/10.1016/j.scib.2021.05.024}{Sci. Bull. {\bf
  66}, 1789}~(2021).

\bibitem{KjaergaardARCMP2020}
Morten Kjaergaard, Mollie~E. Schwartz, Jochen Braumüller, Philip Krantz, Joel
  I.-J. Wang, Simon Gustavsson, and William~D. Oliver.
\newblock ``Superconducting qubits: Current state of play''.
\newblock
  \href{https://dx.doi.org/10.1146/annurev-conmatphys-031119-050605}{Annu. Rev.
  Condens. Matter Phys. {\bf 11}, 369}~(2020).

\bibitem{BlaisRMP2021}
Alexandre Blais, Arne~L. Grimsmo, Steven~M. Girvin, and Andreas Wallraff.
\newblock ``Circuit quantum electrodynamics''.
\newblock \href{https://dx.doi.org/10.1103/revmodphys.93.025005}{Rev. Mod.
  Phys. {\bf 93}, 025005}~(2021).

\bibitem{ChildsPRA2003}
Andrew Childs, Henry Haselgrove, and Michael Nielsen.
\newblock ``Lower bounds on the complexity of simulating quantum gates''.
\newblock \href{https://dx.doi.org/10.1103/PhysRevA.68.052311}{Phys. Rev. A
  {\bf 68}, 052311}~(2003).

\bibitem{weylchamber}
Michael~H. Goerz and {Contributors}~(2022).
\newblock
  code:~\href{https://github.com/qucontrol/weylchamber}{qucontrol/weylchamber}.

\bibitem{QuantumControlBase-jl}
Michael~H. Goerz and {Contributors}~(2022).
\newblock
  code:~\href{https://github.com/JuliaQuantumControl/QuantumControlBase.jl}{JuliaQuantumControl/QuantumControlBase.jl}.

\bibitem{MakhlinQIP2002}
Yuriy Makhlin.
\newblock ``Nonlocal properties of two-qubit gates and mixed states, and the
  optimization of quantum computations''.
\newblock \href{https://dx.doi.org/10.1023/A:1022144002391}{Quantum Inf.
  Process. {\bf 1}, 243}~(2002).

\bibitem{DormandJCAM1980}
John~R. Dormand and Pete~J. Prince.
\newblock ``A family of embedded {Runge}-{Kutta} formulae''.
\newblock \href{https://dx.doi.org/10.1016/0771-050x(80)90013-3}{J. Comput.
  Appl. Math {\bf 6}, 19}~(1980).

\bibitem{RackauckasJORS2017}
Christopher Rackauckas and Qing Nie.
\newblock ``{DifferentialEquations.jl} -- a performant and feature-rich
  ecosystem for solving differential equations in {Julia}''.
\newblock \href{https://dx.doi.org/10.5334/jors.151}{J. Open Res. Softw.{\bf
  5}}~(2017).

\bibitem{BenchmarkTools-jl}
Jarrett Revels and {Contributors}~(2022).
\newblock
  code:~\href{https://github.com/JuliaCI/BenchmarkTools.jl}{JuliaCI/BenchmarkTools.jl}.

\bibitem{LuAPS2022}
Yunwei Lu, Vinh~San Dinh, and Jens Koch.
\newblock ``Increasing memory and runtime performance of {GRAPE} for control in
  large quantum systems''.
\newblock In Bulletin of the American Physical Society, APS March Meeting 2022,
  Chicago.
\newblock Number~10 in Session Y41~(2022).
\newblock
  url:~\href{https://meetings.aps.org/Meeting/MAR22/Session/Y41.10}{meetings.aps.org/Meeting/MAR22/Session/Y41.10}.

\bibitem{psutil}
Giampaolo Rodola and {Contributors}~(2022).
\newblock  code:~\href{https://github.com/giampaolo/psutil}{giampaolo/psutil}.

\bibitem{FiniteDifferences-jl}
Will Tebbutt, Frames~Catherine White, Miha Zgubic, Wessel Bruinsma, and
  {Contributors}~(2022).
\newblock
  code:~\href{https://github.com/JuliaDiff/FiniteDifferences.jl}{JuliaDiff/FiniteDifferences.jl}.

\bibitem{MachnesPRL2018}
Shai Machnes, Elie Assémat, David~J. Tannor, and Frank~K. Wilhelm.
\newblock ``Tunable, flexible, and efficient optimization of control pulses for
  practical qubits''.
\newblock \href{https://dx.doi.org/10.1103/PhysRevLett.120.150401}{Phys. Rev.
  Lett. {\bf 120}, 150401}~(2018).

\bibitem{SorensenPRA2018}
Jens Jakob W.~H. Sørensen, Mikel~O. Aranburu, Till Heinzel, and Jacob~F.
  Sherson.
\newblock ``Quantum optimal control in a chopped basis: Applications in control
  of {Bose-Einstein} condensates''.
\newblock \href{https://dx.doi.org/10.1103/PhysRevA.98.022119}{Phys. Rev. A
  {\bf 98}, 022119}~(2018).

\bibitem{LucarelliPRA2018}
Dennis Lucarelli.
\newblock ``Quantum optimal control via gradient ascent in function space and
  the time-bandwidth quantum speed limit''.
\newblock \href{https://dx.doi.org/10.1103/physreva.97.062346}{Phys. Rev. A
  {\bf 97}, 062346}~(2018).

\bibitem{GoodwinJCP2016}
David~L. Goodwin and Ilya Kuprov.
\newblock ``Modified {Newton}-{Raphson} {GRAPE} methods for optimal control of
  spin systems''.
\newblock \href{https://dx.doi.org/10.1063/1.4949534}{J. Chem. Phys. {\bf 144},
  204107}~(2016).

\bibitem{Diffractor-jl}
Keno Fischer and {Contributors}~(2022).
\newblock
  code:~\href{https://github.com/JuliaDiff/Diffractor.jl}{JuliaDiff/Diffractor.jl}.

\bibitem{ReichPhD2015}
Daniel~M. Reich.
\newblock ``Efficient characterisation and optimal control of open quantum
  systems. {M}athematical foundations and physical applications''.
\newblock PhD thesis.
\newblock Universität Kassel.
\newblock ~(2015).
\newblock
  url:~\href{https://d-nb.info/1073888851/34}{d-nb.info/1073888851/34}.

\bibitem{PlatzerPRL2010}
Felix Platzer, Florian Mintert, and Andreas Buchleitner.
\newblock ``Optimal dynamical control of many-body entanglement''.
\newblock \href{https://dx.doi.org/10.1103/physrevlett.105.020501}{Phys. Rev.
  Lett. {\bf 105}, 020501}~(2010).

\bibitem{CanevaNJP2012}
Tommaso Caneva, Tommaso Calarco, and Simone Montangero.
\newblock ``Entanglement-storage units''.
\newblock \href{https://dx.doi.org/10.1088/1367-2630/14/9/093041}{New J. Phys.
  {\bf 14}, 093041}~(2012).

\bibitem{DoriaPRL2011}
Patrick Doria, Tommaso Calarco, and Simone Montangero.
\newblock ``Optimal control technique for many-body quantum dynamics''.
\newblock \href{https://dx.doi.org/10.1103/PhysRevLett.106.190501}{Phys. Rev.
  Lett. {\bf 106}, 190501}~(2011).

\bibitem{CanevaPRA2011}
Tommaso Caneva, Tommaso Calarco, and Simone Montangero.
\newblock ``Chopped random-basis quantum optimization''.
\newblock \href{https://dx.doi.org/10.1103/PhysRevA.84.022326}{Phys. Rev. A
  {\bf 84}, 022326}~(2011).

\bibitem{RachPRA2015}
Niklas Rach, Matthias~M. Müller, Tommaso Calarco, and Simone Montangero.
\newblock ``Dressing the chopped-random-basis optimization: A bandwidth-limited
  access to the trap-free landscape''.
\newblock \href{https://dx.doi.org/10.1103/PhysRevA.92.062343}{Phys. Rev. A
  {\bf 92}, 062343}~(2015).

\bibitem{LeeS2007}
Hohjai Lee, Yuan-Chung Cheng, and Graham~R. Fleming.
\newblock ``Coherence dynamics in photosynthesis: Protein protection of
  excitonic coherence''.
\newblock \href{https://dx.doi.org/10.1126/science.1142188}{Science {\bf 316},
  1462}~(2007).

\bibitem{HuelgaCP2013}
Susana~F. Huelga and Martin~B. Plenio.
\newblock ``Vibrations, quanta and biology''.
\newblock \href{https://dx.doi.org/10.1080/00405000.2013.829687}{Contemp. Phys.
  {\bf 54}, 181}~(2013).

\bibitem{GoerzSPIEO2021}
Michael~H. Goerz, Mark~A. Kasevich, and Vladimir~S. Malinovsky.
\newblock ``Quantum optimal control for atomic fountain interferometry''.
\newblock In Proc. SPIE 11700, Optical and Quantum Sensing and Precision
  Metrology.
\newblock ~(2021).
\newblock
  url:~\href{https://doi.org/10.1117/12.2587002}{doi.org/10.1117/12.2587002}.

\bibitem{DuspayevPRA2021}
A.~Duspayev and G.~Raithel.
\newblock ``Tractor atom interferometry''.
\newblock \href{https://dx.doi.org/10.1103/physreva.104.013307}{Phys. Rev. A
  {\bf 104}, 013307}~(2021).

\bibitem{RaithelQST2022}
Georg~A. Raithel, Alisher Duspayev, Bineet Dash, Sebastián~C. Carrasco,
  Michael~H. Goerz, Vladan Vuletic, and Vladimir~S. Malinovsky.
\newblock ``Principles of tractor atom interferometry''.
\newblock \href{https://dx.doi.org/10.1088/2058-9565/ac9429}{Quantum Sci.
  Technol.}~(2022).

\bibitem{SekatskiQ2017}
Pavel Sekatski, Michalis Skotiniotis, Janek Kołodyński, and Wolfgang Dür.
\newblock ``Quantum metrology with full and fast quantum control''.
\newblock \href{https://dx.doi.org/10.22331/q-2017-09-06-27}{Quantum {\bf 1},
  27}~(2017).

\bibitem{LinPRA2021}
Chungwei Lin, Yanting Ma, and Dries Sels.
\newblock ``Optimal control for quantum metrology via {Pontryagin}'s
  principle''.
\newblock \href{https://dx.doi.org/10.1103/physreva.103.052607}{Phys. Rev. A
  {\bf 103}, 052607}~(2021).

\bibitem{CarrascoPRA2022}
Sebastián~C. Carrasco, Michael~H. Goerz, Zeyang Li, Simone Colombo, Vladan
  Vuletić, and Vladimir~S. Malinovsky.
\newblock ``Extreme spin squeezing via optimized one-axis twisting and
  rotations''.
\newblock \href{https://dx.doi.org/10.1103/physrevapplied.17.064050}{Phys. Rev.
  Applied {\bf 17}, 064050}~(2022).

\bibitem{semiad_data}
Michael~H. Goerz, Sebastián~C. Carrasco, and Vladimir~S. Malinovsky~(2022).
\newblock
  code:~\href{https://github.com/ARLQCI/2022-04\_semiad\_paper}{ARLQCI/2022-04\_semiad\_paper}.

\end{thebibliography}

\end{document}